# Modélisation factorielle des interactions entre deux ensembles d'observations : la méthode PLS-FILM[1] (*Partial Least Squares Factor Interaction Linear Modelling*)

X. Bry * , T. Verron **

* Département de Mathématiques, UM2, Place Eugène Bataillon, 34090 Montpellier

** ALTADIS, Centre de recherche SCR, 4 rue André Dessaux, 45000 Fleury lès Aubrais

*Résumé:*

*On considère un tableau codant numériquement des interactions entre deux ensembles d'observations respectivement dénommés "sujets" et "objets". Par ailleurs, on dispose de descriptions des sujets et des objets à l'aide de variables. Nous proposons ici une technique d'analyse exploratoire des interactions. Cette technique géométrique utilise une modélisation factorielle hiérarchique des interactions sujets-objets à partir de descriptions structurelles respectives des uns et des autres. Elle fournit un pont entre les méthodes RLQ de Chessel et L-PLS de Martens, avec lesquelles elle n'a cependant de commun que les composantes de rang 1.*

*Mots-clés:*

*Equations Structurelles, Interactions, L-PLS, Modèle Linéaire, PLS, PLS-FILM, Régression Linéaire, RLQ, Variables Latentes.*

*Abstract:*

*In this work, we consider a data array encoding interactions between two sets of observations respectively referred to as "subjects" and "objects". Besides, descriptions of subjects and objects are available through two variable sets. We propose a geometrically grounded exploratory technique to analyze the interactions using descriptions of subjects and objects: interactions are modelled using a hierarchy of subject-factors and object-factors built up from these descriptions. Our method bridges the gap between those of Chessel (RLQ analysis) and Martens (L-PLS), although it only has rank 1 components in common with them.*

*Keywords:*

*Interactions, Latent Variables, Linear Model, Linear Regression, L-PLS, PLS, PLS-FILM, RLQ, Structural Equation Modelling.*

Notations:

Les minuscules carolingiennes désignent en général des vecteurs colonnes (*a,b,...x, y...*), ou des valeurs courantes d'indices (*j, k..., s, t...*).

$<u_1, \ldots , u_n>$ désigne le sous-espace vectoriel engendré par les vecteurs $u_1, \ldots , u_n$.

$e_n$ désigne le vecteur de $\mathbb{R}^n$ dont toutes les composantes valent 1.

Les minuscules grecques ($\alpha, \beta, \ldots \lambda, \mu, \ldots$) désignent en général des scalaires.

1 Dans l'article, on emploiera l'acronyme abrégé: FILM

$<x|y>_M$ désigne le produit scalaire des vecteurs $x$ et $y$ au sens de la métrique $M$.

Les majuscules désignent en général des matrices (*A, B...X, Y...*), ou des valeurs maximales d'indices (*J, K...S, T...*).

$X$ étant une matrice $(I,J)$:

> $x_i^j$ désigne la valeur située à l'intersection de la ligne $i$ et de la colonne $j$;
>
> $x_i$ désigne le vecteur $(x_i^j)_{j=1 \text{ à } J}$ ; $x^j$ désigne le vecteur $(x_i^j)_{i=1 \text{ à } I}$
>
> $<X>$ désigne le sous-espace vectoriel engendré par les colonnes de $X$.

$\Pi_E$ désigne le projecteur orthogonal sur un sous-espace $E$, au sens d'une métrique à préciser. $X$ étant une matrice, on notera de façon allégée $\Pi_X$ le projecteur $\Pi_{<X>}$

Dans un algorithme, $a(k)$ désigne la valeur de l'élément $a$ après l'itération $k$.

Abréviations:

AC*: Analyse Canonique* ACB*: Analyse des Correspondances Binaires*

ACP*: Analyse en Composantes Principales*

ACPVI: *ACP sur Variables Instrumentales* (= *ARM*)

AIB*: Analyse Inter-Batteries*

ARM*: Analyse des Redondances Maximales* (= *ACPVI*)

C.L.*: Combinaison linéaire* L-PLS : *régression PLS pour structures en L.*

OLS: *Ordinary Least Squares* OLS1*: Régression OLS univariée*

PLS: *Partial Least Squares* PLS1*: Régression PLS univariée (1 variable dépendante)*

PLSn*: Régression PLS multivariée (1 groupe de variables dépendantes)*

# Introduction

Dans de nombreux domaines, des interactions apparaissent naturellement entre deux ensembles d'observations, et la question se pose de modéliser ces interactions en fonction des caractéristiques des observations. Les observations des deux ensembles seront respectivement appelées *sujets* et *objets*. Le choix des sujets et des objets est arbitraire, les deux ensembles étant traités de façon parfaitement symétrique.

Par exemple, en analyse sensorielle, on peut faire noter par un jury l'accord d'un vin (sujet) avec un fromage (objet). Vins et fromages sont décrits par ailleurs à l'aide de notes sensorielles et de caractéristiques physico-chimiques. En toute généralité, on peut imaginer toutes sortes de données de correspondance ou d'interaction entre deux ensembles d'individus assez richement décrits par ailleurs:

| Domaine | Sujets | Objets | Interactions |
|---|---|---|---|
| Parfumerie | Base | Parfum | Note d'appréciation/efficacité |
| Cosmétique | Derme | Produit | Note d'appréciation/efficacité |
| Agronomie | Sol | Variété | Performance |
| Ecologie | Environnement | Espèce | Densité de peuplement |
| Médecine | Pathologie | Traitement | Performance (*e.g.* survie) |
| Archéologie | Site | Objets | Densité de présence |
| Marketing | Consommateurs | Produits | Usage (fréquence/quantité) |

Les descriptions des sujets et objets peuvent être très fournies, tandis que les observations elles-mêmes sont en nombre plutôt limité. Il est alors impossible de modéliser immédiatement les interactions en fonction des caractéristiques. En effet, si l'on a $n$ sujets décrits par $J$ variables $x^j$ et $p$ objets décrits par $K$ variables $y^k$, les interactions sont au nombre de $np$, tandis que leur modèle linéaire ou linéaire généralisé ferait intervenir $J+K+JK$ variables (les $x^j$, $y^k$ et $x^jy^k$).

On a dans ce cas besoin d'une méthode exploratoire qui permette de visualiser, dans les descriptions des observations, les dimensions utiles à la modélisation des interactions. La méthode que nous proposons ici, fonctionnant sur le principe de la régression PLS, s'attache en outre à rechercher des dimensions structurellement fortes. Plusieurs auteurs ont proposé des méthodes traitant ce problème. Nous les examinerons rapidement afin d'en montrer les points communs, les différences, et les limitations.

# 1. Données, problème, modèle

## *1.1. Données*

On dispose de 3 tableaux matriciels:

- $X$ est une matrice $(n,J)$ décrivant $n$ observations appelés *sujets* à l'aide de $J$ variables numériques $x^1, \ldots , x^J$. La valeur de $x^j$ pour le sujet $i$ est notée $x_i^j$.
- $Y$ est une matrice $(p,K)$ décrivant $p$ observations appelés *objets* à l'aide de $K$ variables numériques $y^1, \ldots , y^K$. La valeur de $y^k$ pour l'objet $m$ est notée $y_m^k$.
- $Z$ est une matrice $(n,p)$ codant les interactions entre les sujets et les objets à l'aide de valeurs numériques. La valeur de l'interaction entre le sujet $i$ et l'objet $m$ est notée $z_i^m$.

En outre, les sujets (respectivement objets) sont munis de pondérations. Le sujet $i$ est muni du poids statistique $p_i$. On impose: $\sum_{i=1}^{n} p_i = 1$ . De même, L'objet $m$ est muni du poids statistique $q_m$. On impose: $\sum_{m=1}^{p} q_m = 1$ . La matrice $(n,n)$ diagonale des $p_i$ est notée $P$; La matrice $(p,p)$ diagonale des $q_m$ est notée $Q$.

## *1.2. Problème*

On cherche à analyser les interactions en les mettant en rapport avec les caractéristiques des sujets et des objets respectivement. Cette analyse utilisera une modélisation factorielle des interactions sujet-objet: on cherchera, dans les caractéristiques $X$ des sujets (respectivement $Y$ des objets), un petit nombre de composantes assez fortes $f^1,...,f^S$ (resp. $g^1,..., g^T$) dont les interactions puissent rendre compte de la plus grande part possible du tableau $Z$ (cf. figure 1).

## *1.3. Modèle de Z*

Nous allons présenter deux modèles de $Z$. Le premier (modèle A) ne prend en compte que les interactions entre composantes sujet et objet. Le second, plus "réaliste" (modèle B) prend en compte, en sus des interactions, les effets marginaux de ces composantes. Dans un premier temps, seul le modèle A sera estimé, car il mène à des développements formels simplifiés (section 2). Dans un deuxième temps, nous présenterons en détail le modèle B et l'algorithme qui l'estime (section 3).

### a) Espaces métriques

• L'espace $\mathbb{R}^J$ où se trouvent les vecteurs $x_i$ correspondant aux lignes de $X$ est muni d'une métrique euclidienne $M$ fournissant le produit scalaire:

$$\langle x_i | x_{i'} \rangle_M = x_i ' M x_{i'}$$

• L'espace $\mathbb{R}^K$ où se trouvent les vecteurs $y_m$ correspondant aux lignes de $Y$ est muni d'une métrique euclidienne $N$ fournissant le produit scalaire:

$$\langle y_m | y_{m'} \rangle_N = y_m ' N y_{m'}$$

• Les espaces $\mathbb{R}^n$ et $\mathbb{R}^p$ où se trouvent respectivement les variables sujet $x^j$ et objet $y^k$ sont munis des métriques respectives $P$ et $Q$, de sorte que si les variables sont centrées, leur produit scalaire égale leur covariance.

$$\langle x^j | x^l \rangle_P = x^j ' P x^l \quad ; \quad \langle y^k | y^r \rangle_Q = y^k ' Q y^r$$

• L'espace $\mathbb{R}^{np}$ où se trouve le tableau $Z$ est muni du produit scalaire suivant (noté $R$):

$$\langle W | V \rangle_R = tr(QW ' PV)$$

(On vérifie aisément symétrie, bilinéarité et le fait que la forme quadratique associée est définie positive).

N.B. Les métriques $M$ et $N$ offrent une possibilité d'adapter la méthode à des données de structure ou type particulier (*e.g.* données qualitatives, ou structurées en sous-tableaux).

### b) Modèles de *Z*

***• Composantes sujet et objet:***

$Z$ sera modélisé à partir de $X$ et $Y$ en utilisant:

- $S$ composantes-sujet $f^s \in \mathbb{R}^n$ décrivant les sujets et supposées exprimables linéairement à partir de $X$ ;

- $T$ composantes-objet $g^t \in \mathbb{R}^p$ décrivant les objets et supposées exprimables linéairement à partir de $Y$ (cf. figure 1).

*Figure 1: Résumer X et Y pour modéliser Z*

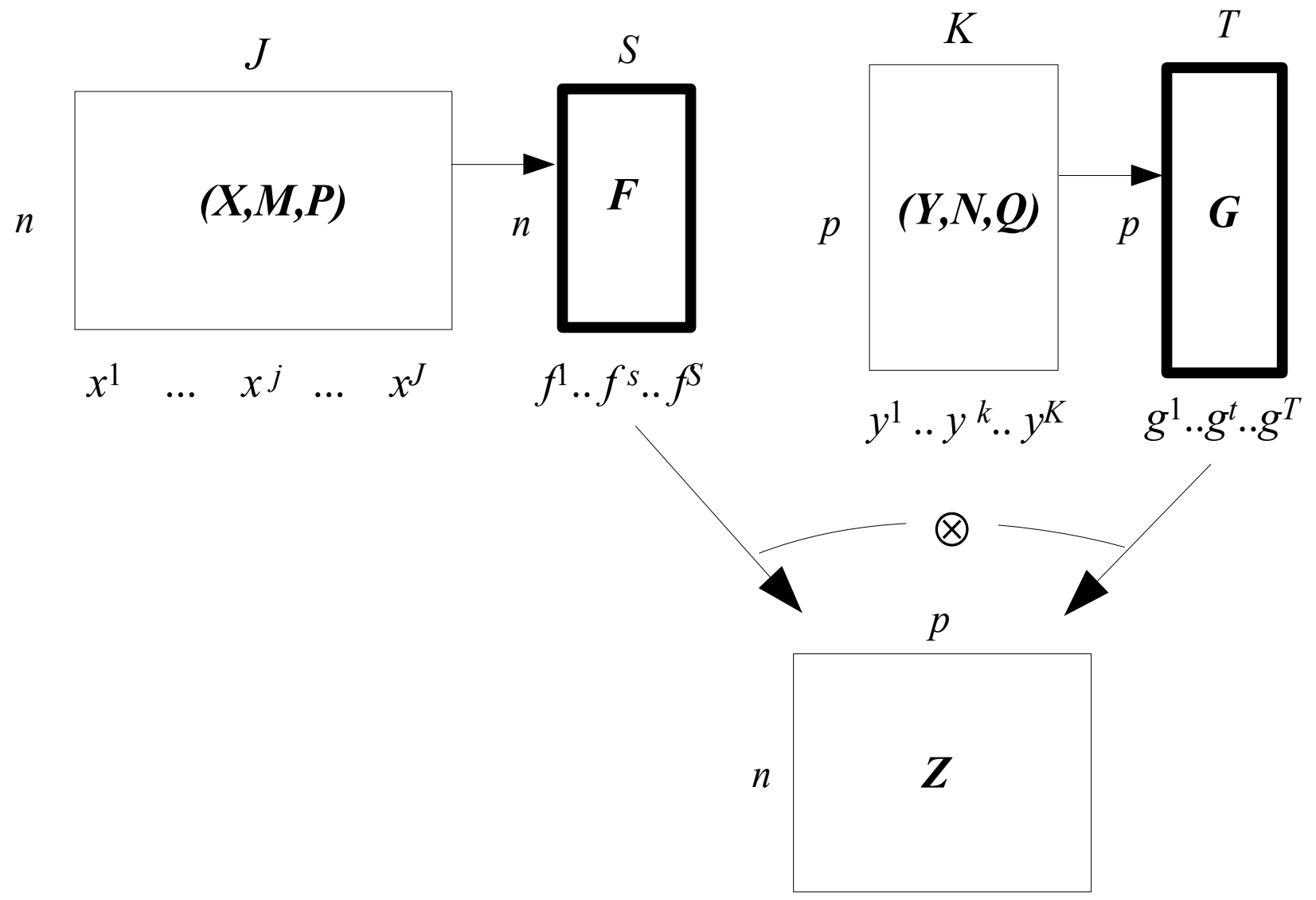

On pose ainsi:

$$\forall i,s: \quad f_i^s = \sum_j a_j^s x_i^j \quad ; \quad \forall m,t: \quad g_m^t = \sum_k b_k^t y_m^k$$

Soit, matriciellement:

$$F = XA \;\; ; \;\; G = YB$$

$$\text{où:} \quad F = (f_i^s)_{i,s} \;\; ; \;\; G = (g_m^t)_{m,t} \;\; ; \;\; A = (a_j^s)_{j,s} \;\; ; \;\; B = (b_k^t)_{k,t}$$

## • *Modèles de Z:*

De façon générale, on cherche à décomposer $Z$ comme suit:

$$Z = \hat{Z} + E$$

où la partie prédite $\hat{Z}$ est fonction des composantes $F$ et $G$, et où $E$ représente une partie résiduelle à minimiser. Selon la formulation de $\hat{Z}$ , on obtient différents modèles.

**Modèle A (interactions pures):**

On formule comme suit $Z$ en fonction des composantes:

$$\forall i,m: \quad \hat{z}_i^m = \sum_{s,t} \omega_{st} f_i^s g_m^t$$

Soit, matriciellement:

$$\hat{Z} = (\hat{z}_i^m)_{i,m} = \sum_{s,t} \omega_{st} f^s g^{t\,\prime} \;=\; F\,\Omega\,G' \quad \text{où:} \quad \Omega = \big((\omega_{st})\big)_{\substack{s=1\,\text{à}\,S\\ t=1\,\text{à}\,T}}$$

$E$ est une matrice de résidus que l'on cherche à minimiser.

L'estimation du modèle A par FILM utilisera un algorithme noté FILM-A.

N.B.: On peut écrire:

$$Z = F\Omega G' + E = XCY' + E \quad \text{avec:} \quad C = A\Omega B'$$

**Modèle B (interactions et effets propres):**

Ici, on tient compte des effets propres des composantes. On pose ainsi:

$$\hat{z}_i^m = \sum_{s,t} \omega_{st} f_i^s g_m^t + \sum_s \phi_s f_i^s + \sum_t \gamma_t g_m^t$$

On obtient alors:

$$\hat{Z} = F\Omega G' + F\Phi e_p' + e_n \Gamma' G' \quad \text{où:} \quad \Phi = (\phi_s)_{s=1 \text{ à } S} \;;\; \Gamma = (\gamma_t)_{t=1 \text{ à } T}$$

L'estimation du modèle B par FILM utilisera un algorithme noté FILM-B.

Notons qu'en général, le tableau $Z$ peut et doit être modélisé à l'aide du modèle B, les effets propres étant susceptibles d'être primordiaux. Le cas particulier extrême est celui d'un modèle linéaire de $Z$ en fonction des $f$ et $g$, où seuls les effets propres de ces dernières comptent.

- ***Orthogonalités:***

Pour des raisons d'économie d'information et pour faciliter les représentations graphiques de chacun des groupes $X$ et $Y$, on demandera aux $f^s$ d'être orthogonales dans $\mathbb{R}^n$ comme aux $g^t$ de l'être dans $\mathbb{R}^p$. Pour des raisons d'identifiabilité, on pourra, *in fine*, normer ces composantes.

L'ajustement du modèle de $Z$ conduit par ailleurs à:

$$\hat{Z} \perp E$$

- ***Interprétation des modèles:***

**Modèle B:**

L'orthogonalité deux à deux des composantes sujet (resp. objet) facilite *a priori* le fait d'imaginer que l'une varie les autres restant fixes. Dès lors:

$$\text{(a)} \quad \Delta f_i^s = 1, \Delta f_i^u = 0 \, \forall u \neq s \quad \Rightarrow \quad \Delta \hat{z}_i^m = \phi_s + \sum_t \omega_{st} g_m^t$$

$$\text{(b)} \quad \Delta g_m^t = 1, \Delta g_m^v = 0 \, \forall v \neq t \quad \Rightarrow \quad \Delta \hat{z}_i^m = \gamma_t + \sum_s \omega_{st} f_i^s$$

Selon (a), si pour le sujet $i$, $f^s$ augmente d'une unité, les autres composantes sujet $f^u$ restant inchangées, l'impact sur l'interaction prévisible entre le sujet $i$ et l'objet $m$ sera la somme de plusieurs variations: une variation "propre" fixe $\phi_s$, et plusieurs variations d'interaction $\omega_{st} g_m^t$ dont chacune est liée à la valeur pour l'objet $m$ d'une composante objet $g^t$.

Symétriquement, selon (b), si pour l'objet $m$, $g^t$ augmente d'une unité, les autres composantes objet $g^v$ restant inchangées, l'impact sur l'interaction prévisible entre le sujet $i$ et l'objet $m$ sera la somme de plusieurs variations: une variation "propre" fixe $\gamma_t$, et plusieurs variations d'interaction $\omega_{st} f_i^s$ dont chacune est liée à la valeur pour le sujet $i$ d'une composante sujet $f^s$.

Pour interpréter le modèle, il faudra donc interpréter les composantes $f$ et $g$ et examiner les « effets » $\omega_{st} g_m^t$ et $\omega_{st} f_i^s$ .

**Modèle A:**

Le modèle A est le cas particulier du modèle B sans effets propres des composantes.

**c) Force structurelle des composantes**

L'estimation du modèle ci-dessus prendra naturellement en compte l'ajustement du modèle aux données (petitesse du résidu $E$). Mais on souhaite également tenir compte d'un critère de force structurelle des composantes, afin de bannir du modèle estimé les dimensions de bruit de chacun des groupes $X$ et $Y$.

Pour mesurer la force structurelle d'une composante, posons:

$$f^s = XMu_s \quad \text{où:} \quad u_s \in \mathbb{R}^J, \|u_s\|_M^2 = 1$$

$\|f^s\|_P^2 = u_s' MX' PXMu_s$ représente ainsi l'inertie des $n$ sujets le long de l'axe $<u_s>$ de $\mathbb{R}^J$ . Ceci nous permet de considérer dans ce cas $\|f^s\|_P$ comme une mesure possible de la force structurelle de la composante $f^s$.

On procédera de même pour les composantes-objet $g^t = YNv_t$: lorsque l'on impose $v_t'Nv_t = 1$, $\|g^t\|_Q^2$ est l'inertie des $p$ objets sur l'axe $<v_t>$ de $\mathbb{R}^K$ . La force structurelle de $g^t$ dans son groupe sera alors mesurée par $\|g^t\|_Q$ .

L'estimation du modèle se fera en optimisant un critère amalgamant qualité d'ajustement et force structurelle des composantes.

**d) Suite ordonnée de modèles**

La méthode FILM construit une suite de modèles ordonnés par inclusion, de la façon suivante:

Le modèle de rang 1 sera obtenu pour $S = T = 1$. Les composantes $f^1$ et $g^1$ seront obtenues par optimisation d'un critère amalgamant force structurelle des composantes et ajustement de $Z$ par un vecteur colinéaire à $f^1 g^{1\prime}$.

Le modèle de rang 2 sera obtenu à partir du modèle de rang 1, en cherchant les composantes $f^2$ et $g^2$ orthogonales respectivement à $f^1$ et $g^1$ maximisant un critère amalgamant force structurelle des composantes et ajustement de $Z$ par une combinaison linéaire des vecteurs: $f^1 g^{1\prime}, f^1 g^{2\prime}, f^2 g^{1\prime}, f^2 g^{2\prime}$. On notera la présence, dès le rang 2, des interactions "croisées" $f^s g^{t\prime}$, $s \neq t$.

De façon générale, Le modèle de rang $t$ sera obtenu à partir du modèle de rang $t$-1, en cherchant les composantes $f^t$ et $g^t$ orthogonales respectivement à $f^1, \ldots f^{t-1}$ et $g^1, \ldots g^{t-1}$ maximisant un critère amalgamant force structurelle des composantes et ajustement de $Z$ par une combinaison linéaire des vecteurs: $f^1 g^{1\prime}, \ldots f^1 g^{t\prime}, \ldots, f^t g^{1\prime}, \ldots f^t g^{t\prime}$.

## 1.4. Dans la littérature

Les auteurs ayant traité de problèmes similaires ont fondé leurs méthodes sur des versions simplifiées du modèle de base présenté ci-dessus. Nous situons brièvement ci-après leurs techniques dans le cadre de ce modèle, afin de mettre en lumière leurs points communs, différences et limitations.

### a) L'analyse RLQ de Chessel

Chessel, dans le sillage de ses travaux sur la co-inertie, a proposé une méthode particulière d'analyse d'un couple de tableaux quelconques (ici $X$ et $Y$) liés par un tableau de contingence (ici $Z$): la méthode RLQ [Chessel *et al.* 1993]. Nous verrons dans la section 3.3. que le même problème peut être traité par FILM. Le modèle de base de RLQ est le modèle A. Cette méthode est fondée sur la maximisation d'un critère de co-inertie qui correspond exactement au critère de covariance de FILM dans le cas des tableaux de contingence. Les méthodes RLQ et FILM-A coïncident alors au rang 1. Cependant, elles fournissent des composantes distinctes à partir du rang 2, pour les raisons suivantes:

1) Les problématiques des deux méthodes diffèrent. L'analyse RLQ se propose d'analyser la co-structure des tableaux $X$ et $Y$, comme l'analyse de co-inertie en général, mais les observations de ces deux tableaux (sujets et objets) étant dans ce cas précis liés par $Z$. L'analyse RLQ est ainsi focalisée sur les structures d'inertie des sujets et des objets. Par suite, elle aboutit à une "simple" diagonalisation. De son côté, FILM cherche bien des structures de $X$ et $Y$, mais seulement dans la mesure où elles jouent un rôle utile dans un modèle de $Z$. FILM est donc davantage focalisée sur la modélisation de $Z$. FILM tire alors parti, dès le rang 2, des interactions croisées entre les composantes (*e.g.*, au rang 2, les interactions entre $f^1$ et $g^2$, ainsi qu'entre $f^2$ et $g^1$), ce que RLQ ne fait pas. Le fait de tenir compte de ces interactions croisées complexifie l'algorithme fournissant les composantes.

En quelque sorte, dans le cas d'un tableau $Z$ de contingence, FILM est à RLQ ce que la régression PLSn est à l'Analyse Inter-Batteries.

2) Plus généralement, l'analyse RLQ suppose que $Z$ est un tableau de contingence. En effet, le modèle A ne convient qu'à un tableau $Z$ doublement centré (en ligne et en colonne); or, comme nous l'exposons en section 3.3.b, le double centrage d'un tableau de contingence fournit un codage adéquat de la liaison entre les deux caractères qu'il croise. Si FILM-A (§2) ne fonctionne de même qu'avec un tableau $Z$ doublement centré (§3.1), il est ensuite étendu par FILM-B au cas d'un tableau $Z$ quelconque (§3.2).

### b) ACP du tableau *Z* avec information sur les marges (ACPIM)

Par ailleurs, Takane et Shibayama ont proposé [Takane et al. 1991] une technique de décomposition d'un tableau $Z$ quelconque, analogue à celle de l'ACP, mais dont les composantes principales sujet $f^s$ (respectivement objet $g^t$) sont astreintes à être linéairement exprimables en fonction des caractéristiques $x^j$ (resp. $y^k$). Le modèle qu'ils posent est:

$$Z = XCY' + XH + KY' + E$$

On constate qu'il n'est pas fait ici de référence *explicite* aux composantes. En particulier, la modélisation des interactions et celle des effets propres n'est pas contrainte à utiliser les mêmes composantes. Dans cette modélisation, seules les structures de $Z$ sont prises en compte, et la force structurelle des C.L. de $X$ (resp. de $Y$) utilisées par la modélisation n'intervient pas. Ainsi, les dimensions résiduelles (bruit) de $X$ et $Y$ peuvent participer à la modélisation de $Z$ au même titre que leurs dimensions fortes. Il s'ensuit une perte de robustesse des composantes, ainsi que des difficultés d'interprétation, notamment en cas de colinéarités dans $X$ et $Y$.

La méthode FILM que nous proposons est, comme l'ACPIM, une méthode d'analyse d'un tableau $Z$ quelconque, qu'elle cherche à modéliser à partir de l'information disponible sur ses marges. Mais à la différence de l'ACPIM, elle cherche à appuyer le modèle de $Z$ sur les structures les plus fortes possible de $X$ et $Y$, afin de fournir un modèle plus robuste et facilement interprétable.

En quelque sorte, FILM est à l'ACPIM ce que la régression PLS1 est à la régression multiple.

### c) L-PLS:

[Martens *et al.* 2005] utilise au départ le modèle A avec des matrices $M, N, P, Q$ toutes égales à l'identité. Pour estimer le modèle, il propose une technique, L-PLS, fondée sur la décomposition en valeurs singulières de la matrice $X'ZY$. Deux variantes sont proposées. Dans la première, les composantes sont: $f^k = Xu_k$ et $g^k = Yv_k$ , où $u_k$ et $v_k$ sont les vecteurs propres normés des matrices respectives $X'ZYY'Z'X$ et $Y'Z'XX'ZY$, associés à la $k^{\text{ième}}$ valeur propre par ordre décroissant. Cette première variante est très semblable à RLQ, avec un système de poids différent.

Dans la seconde variante, les composantes de rang 1 sont les mêmes que dans la première, mais les composantes de rangs 2 sont obtenues en réitérant le procédé donnant les composantes de rang 1 après avoir remplacé les tableaux $X$ et $Y$ par leurs résidus de régression respectivement sur $f^1$ et $g^1$ (il est dès lors évident que seule l'interaction entre $f^2$ et $g^2$ est prise en compte). Ainsi de suite pour les composantes de rangs ultérieurs. Une fois obtenues les composantes, la matrice $\Omega$ est estimée par ajustement à $Z$ du modèle A selon un programme de moindres carrés ordinaires, lequel fournit:

$$\hat{\Omega}=(F'F)^{-1}F'ZG(G'G)^{-1}$$

La diagonalisation initiale n'est pas explicitement présentée comme résultant de l'optimisation d'un critère. Nous montrons toutefois, dans le §2.1.*b*, que tel est le cas, et que ce critère est d'interprétation statistique directe. La méthode FILM-A coïncide avec L-PLS au rang 1, dans le cas particulier de données numériques et d'observations équipondérées. Mais elle s'en distingue dès le rang 2 pour deux raisons essentielles:

1) Comme RLQ, L-PLS (quelle qu'en soit la variante) ne calcule pas les composantes sujet $f$ (resp. objet $g$) de rang $k$ en tenant compte de leurs possibles interactions avec les composantes objet $g$ (resp. sujet $f$) de rangs inférieurs. Il s'ensuit une déperdition éventuellement forte du pouvoir prédictif des composantes.

2) L-PLS calcule les composantes sujet et objet en fonction du pouvoir prédictif de leurs seules interactions, ce qui correspond à l'ajustement du modèle A (encore ces interactions sont-elles ici restreintes aux composantes de même rang). Une fois calculées les composantes $f$ et $g$, L-PLS procède, *via* une régression OLS, à un ajustement à $Z$ du modèle B, ce qui n'est guère cohérent. Par contraste, FILM-B procède au calcul des composantes et à l'ajustement sur le *même* modèle: B.

### d) 2-step L-PLS

Dans [Esposito-Vinzi *et al.* 2007], il est dit que la version L-PLS de Martens ne dérive pas de l'optimisation d'un critère statistiquement interprétable. De fait, aucun critère n'est présenté par Martens. Il n'en demeure pas moins que sa méthode découle au rang 1, comme nous le montrons au §2.1.*b*, de la maximisation d'un critère de covariance étendu au modèle A.

[Esposito-Vinzi *et al.* 2007] proposent, pour remédier à ce manque apparent de critère, une variante de L-PLS qui estime *successivement* les composantes $f$, puis $g$, à l'aide de deux régressions PLS classiques emboîtées. Il y a donc maximisation successive de deux critères partiels. Nous pensons que cette rupture de symétrie dans le traitement des sujets et objets est dommageable, ces deux groupes d'observations jouant des rôles parfaitement symétriques dans le modèle.

**e) Bilan synoptique**

| *Méthode:* <br> *Caractéristiques* | **RLQ** | **ACP-IM** | **L-PLS** | **2-step LPLS** | **FILM** |
|---|---|---|---|---|---|
| Tableau $Z$ | Tableau de Contingence | Variables numériques | Variables numériques | Variables numériques | Quelconque |
| Tableau(x) analysés | $X, Y$ | $Z$ | $Z$ | $Z$ | $Z$ |
| Structures prises en compte | $X, Y, Z$ | $Z$ | $X, Y, Z$ | $X, Y, Z$ | $X, Y, Z$ |
| Modèle utilisé pour l'estimation des composantes | *A* | *B* composantes distinctes pour marges et interactions | *A* | *A* | *A,B* composantes identiques ou distinctes pour marges et interactions |
| Mention d'un critère *global* à optimiser | Oui (co-inertie) | Oui ($R^2$) | Non | Non | Oui (Covariance) |
| Suite de modèles emboîtés | Oui | Non | Oui | Oui | Oui |
| Pondérations quelconques des sujets et objets | Non | Non | Non | Non | Oui |
| Estimations des composantes sujets et objets | Simultanées | Simultanées | Simultanées | Successives | Simultanées |
| Prise en compte des interactions croisées entre composantes lors de l'estimation des composantes | Non | Non | Non | Non | Oui |

# 2. Ajustement de *Z* au modèle A: programme de base, solution, propriétés, algorithme FILM-A

## *2.1. Programme d'ajustement*

### a) Préliminaires

#### • *Un cône*

Considérons le sous-ensemble $C_1$ de $\mathbf{R}^{np}$ défini par:

$$C_1 = \{ Xa(Yb)' | a \in \mathbb{R}^J, b \in \mathbb{R}^K \}$$

$C_1$ n'est pas un espace vectoriel. En effet, il n'est pas stable par addition. Par contre, c'est un cône, étant stable par multiplication par un scalaire:

$$S \in C_1 \Rightarrow \exists a, b : S = Xa(Yb)' \quad ; \quad \lambda \in \mathbb{R} \Rightarrow \lambda S = X \lambda a (Yb)' \in C_1$$

Les coupes de $C_1$ à $a$ (respectivement $b$) fixé sont, elles, des sous-espaces vectoriels.

- ***Correspondance entre produits scalaires dans*** $\mathbb{R}^J, \mathbb{R}^K, C_1$ **:**

$\forall f, h \in \mathbb{R}^J, \forall g, l \in \mathbb{R}^K$:

$$\langle f g' | h l' \rangle_R = tr(Q(fg')' P(hl')) = tr(Qgf'Phl') = tr(l'Qgf'Ph) = (f'Ph)(l'Qg)$$

$$\Leftrightarrow \quad \langle f g' | h l' \rangle_R = \langle f | h \rangle_P \langle g | l \rangle_Q \quad (1)$$

En particulier:

$$\| f g' \|_R^2 = \| f \|_P^2 \| g \|_Q^2 \quad (2)$$

Il découle de cela une correspondance entre vecteurs unitaires de $\mathbb{R}^J, \mathbb{R}^K, C_1$ :

<u>Théorème 1</u>:

(i) Soit $U \in C_1$ : $U = fg'$, où $f \in \mathbb{R}^J, g \in \mathbb{R}^K$. On a:

$$\| f \|_P = 1 \text{ et } \| g \|_Q = 1 \Rightarrow \| U \|_R = 1$$

(ii) Réciproquement, soit un vecteur unitaire $U$ de $C_1$, on peut trouver $f \in \mathbb{R}^J, g \in \mathbb{R}^K$ unitaires tels que $U = fg'$.

<u>Preuve</u>:

(i) découle directement de (2).

(ii):

$$U \in C_1 \Rightarrow \exists F \in \mathbb{R}^J, G \in \mathbb{R}^K / U = FG' \quad ; \quad \| U \|_R^2 = 1 \Rightarrow \| F \|_P^2 \| G \|_Q^2 = 1 \quad \Leftrightarrow \quad \| F \|_P = \frac{1}{\| G \|_Q}$$

Donc:

$$U = \left( \frac{F}{\| F \|_P} \right) \left( \frac{G}{\| G \|_Q} \right)'$$

■

Il découle également de (1) une correspondance entre vecteurs orthogonaux de $\mathbb{R}^J, \mathbb{R}^K, C_1$ :

<u>Théorème 2</u>:

Soient $U = fg'$ et $V = f^* g^{*\prime}$ deux vecteurs de $C_1$. On a immédiatement, d'après (1):

$$U \perp V \Leftrightarrow f \perp f^* \text{ ou } g \perp g^* \quad (3)$$

**• *Ajustement d'un vecteur par un vecteur d'un cône:***

Soit $L$ un espace métrique quelconque, soit $Z \in L, \|Z\|^2 = 1$ et soit $C$ un cône de $L$. Nous allons chercher la meilleure approximation de $Z$ appartenant à $C$. Le fait de prendre $Z$ normé simplifie les écritures, mais n'est aucunement limitatif, le problème de l'ajustement sur un cône étant auto-homothétique.

Théorème 3:

Soit:

$$\tilde{Z} = Arg\, min_{S \in C} \|S - Z\|^2$$

Alors:

$$\tilde{Z} = \langle U | Z \rangle U \quad \text{où:} \quad U = Arg\, Max_{U \in C, \|U\|^2 = 1} \cos(U, Z) \quad (4)$$

Preuve:

Utilisons le fait que $C$ est un cône pour poser: $S = \alpha U$ où $\alpha \in \mathbb{R}^+$ et $U \in C, \|U\|^2 = 1$ . On cherche donc à résoudre:

$$\underset{\substack{\alpha \in \mathbb{R}^+ \\ U \in C, \|U\|^2 = 1}}{Min} \|\alpha U - Z\|^2$$

Or:

$$\|\alpha U - Z\|^2 = \alpha^2 - 2\alpha \langle U | Z \rangle + \|Z\|^2$$

La minimisation sur $\alpha$ conduit à la condition du premier ordre:

$$\left( \frac{\partial}{\partial \alpha} \|\alpha U - Z\|^2 \right)_{\hat{\alpha}} = 0 \quad \Leftrightarrow \quad \hat{\alpha} = \langle U | Z \rangle$$

N.B.: On peut toujours prendre $\hat{\alpha}$ positif, quitte à remplacer $U$ par $-U$.

On a alors:

$$\|\hat{\alpha} U - Z\|^2 = \|Z\|^2 - \langle U | Z \rangle^2 = 1 - \langle U | Z \rangle^2$$

Donc, compte tenu de la positivité de $<U|Z>$, on peut écrire:

$$Min_{U \in C, \|U\|^2 = 1} \|\hat{\alpha} U - Z\|^2 \Leftrightarrow Max_{U \in C, \|U\|^2 = 1} \langle U | Z \rangle \Leftrightarrow Max_{U \in C, \|U\|^2 = 1} \cos(U, Z)$$

■

**b) Le programme d'ajustement de *Z* au modèle A**

• Les critères d'inertie utilisés pour mesurer la force structurelle des composantes et le critère d'ajustement au cône $C_1$ sont simplement amalgamés dans le programme suivant:

$$\boldsymbol{P}(Z; (X,M),(Y,N)): \quad Max_{\substack{u'Mu = 1 \\ v'Nv = 1}} \langle Z | XMu (YNv)' \rangle_R$$

• Justification heuristique du critère maximisé:

Notons: $f = XMu$ , $g = YNv$.

$$\langle Z | fg' \rangle_R = \|fg'\|_R \|Z\|_R \cos_R(fg', Z)$$

$$(2) \Rightarrow \|fg'\|_R = \|f\|_P \|g\|_Q$$

$$\Rightarrow \langle Z | XMu(YNv)' \rangle_R = \|XMu\|_P \|YNv\|_Q \|Z\|_R \cos_R(fg', Z)$$

Ce critère fait ainsi apparaître un produit de trois facteurs dont la maximisation isolée a une interprétation claire:

- $Max_{u'Mu=1} \|XMu\|_P$ correspond à la recherche de composantes fortes dans *X* (ce programme conduit à l'ACP de *X,M,P*).
- $Max_{v'Nv=1} \|YNv\|_Q$ correspond à la recherche de composantes fortes dans *Y* (ce programme conduit à l'ACP de *Y,N,Q*).
- $Max_{f \in \mathbb{R}^n, g \in \mathbb{R}^p} \cos_R(fg', Z)$ correspond, d'après (4), à la recherche du vecteur de $C_1$ le plus proche de *Z*.

*Figure 2: Schéma géométrique du problème*

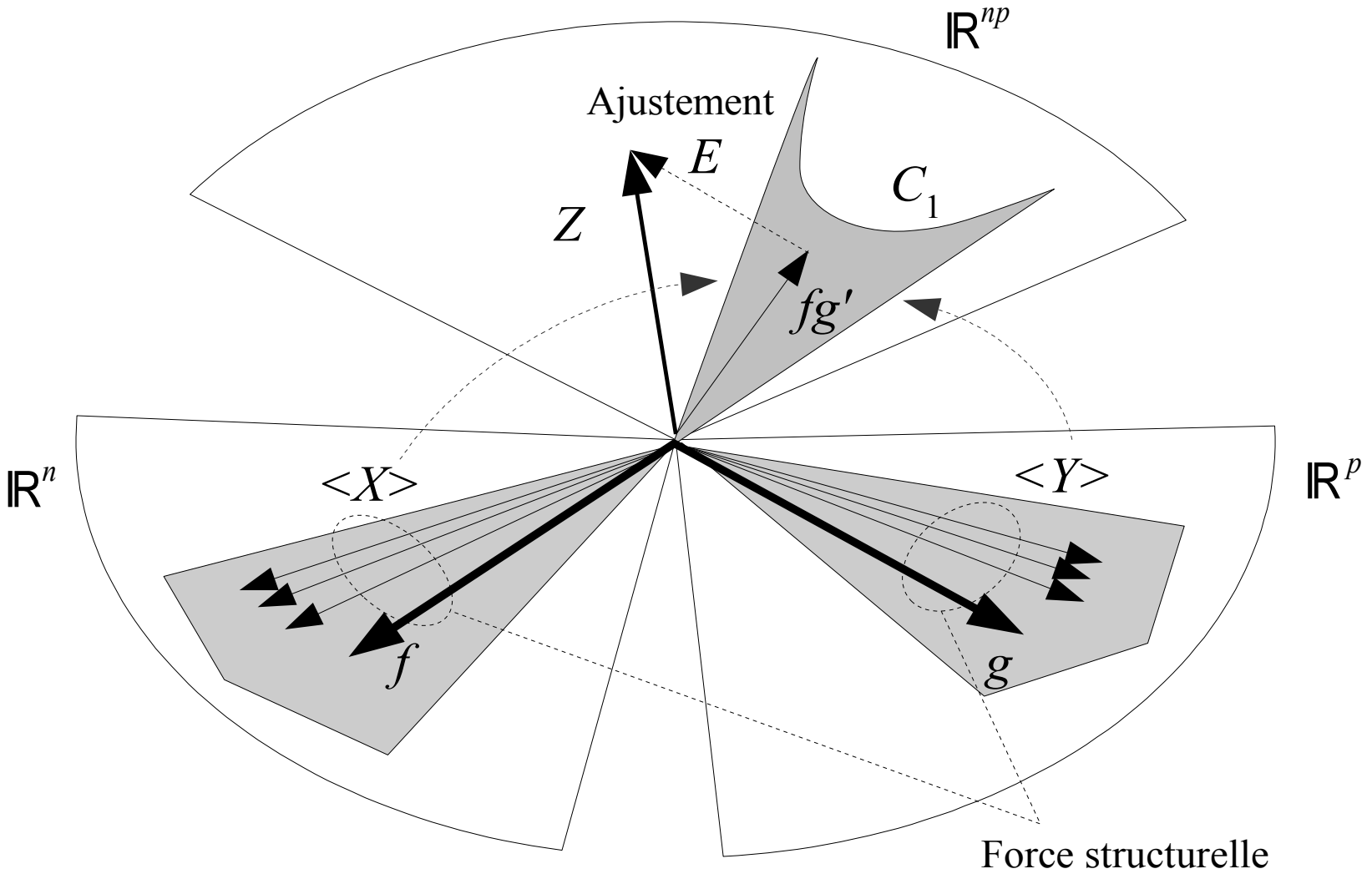

## *2.2. Résolution du programme* **P** *au rang 1*

### a) Vecteurs et composantes

On récrit de façon plus commode le critère à maximiser:

$$\langle Z | XMu(YNv)' \rangle_R = tr(QZ'PXMuv'N'Y') = tr(v'N'Y'QZ'PXMu)$$

$$= v'NY'QZ'PXMu \in \mathbb{R} \quad (5)$$

Le lagrangien du programme est:

$$L = v'NY'QZ'PXMu - \lambda(u'Mu - 1) - \mu(v'Nv - 1)$$

$$\frac{\partial L}{\partial u} = 0 \Leftrightarrow MX'PZQYNv = 2\lambda Mu \quad (6a)$$

$$\frac{\partial L}{\partial v}=0 \Leftrightarrow NY'QZ'PXMu=2\mu Nv \quad (6b)$$

On a: $u'(6a) = v(6b)' = v'NY'QZ'PXMu = 2\lambda = 2\mu$, ce qui implique: $\lambda = \mu$. On posera:

$$\eta=4\lambda^2=4\mu^2$$

Comme le programme maximise $v'NY'QZ'PXMu$, on devra avoir $\eta$ maximum.

Posons $R_{X,M} = XMX'$ et $R_{Y,N} = YNY'$. Les équations (6a) et (6b) entraînent:

$$X(6a) \Leftrightarrow \quad XMX'PZQYNv=\sqrt{\eta}\,XMu \quad \Leftrightarrow \quad R_{X,M}PZQg=\sqrt{\eta}\,f \quad (7a)$$

$$Y(6b) \Leftrightarrow \quad YNY'QZ'PXMu=\sqrt{\eta}\,YNv \quad \Leftrightarrow \quad R_{Y,N}QZ'Pf=\sqrt{\eta}\,g \quad (7b)$$

On en déduit les équations correspondantes dont les composantes $f$ et $g$ sont solutions:

$$R_{X,M}(PZQ)R_{Y,N}(QZ'P)f=\eta f \quad (8a)$$

$$R_{Y,N}(QZ'P)R_{X,M}(PZQ)g=\eta g \quad (8b)$$

On notera $f^1$ et $g^1$ les solutions correspondant à la plus grande valeur propre $\eta_1$.

N.B. Soit $T$ le rang commun aux matrices $R_{X,M}PZQR_{Y,N}QZ'P$ et $R_{Y,N}QZ'PR_{X,M}PZQ$ du système (8). La diagonalisation de ces matrices conduit à $T$ triplets ($f^t$, $g^t$, $\eta^t$) ordonnés par valeur propre $\eta^t$ décroissante. Ces triplets sont ceux de l'analyse RLQ. Ce sont également, lorsque $M$, $N$, $P$, $Q$ sont des matrices identité, ceux de la première variante de L-PLS, consistant à diagonaliser $X'ZYY'Z'X$ et $Y'Z'XX'ZY$, ce à quoi conduisent (6a) et (6b). Ces triplets n'ont en général pas de raison de fournir la meilleure modélisation de type A de $Z$. Dans cette dernière, en effet, interviennent les interactions croisées (*i.e.* entre $f^s$ et $g^t$, $t \neq s$), qui ne sont pas prises en compte dans le critère maximisé dans ***P***. On ne pourrait avoir coïncidence entre les deux analyses que si les interactions croisées ne jouaient aucun rôle dans la modélisation de $Z$ (cf §§ 2.2.*c* et 2.3.*b* ci-dessous).

### b) Equations caractéristiques alternatives

(7a) $R_{X,M}PZQg=\sqrt{\eta}\,f \quad \Rightarrow \quad R_{X,M}PZQgg'QZ'PR_{X,M}Pf=\eta f(f'Pf)$

De même:

(7b) $R_{Y,N}QZ'Pf=\sqrt{\eta}\,g \quad \Rightarrow \quad R_{Y,N}QZ'Pff'PZQR_{Y,N}Qg=\eta g(g'Qg)$

Si l'on ne s'intéresse qu'à déterminer les directions des composantes, on peut conserver ces équations en $P$-normant $f$ et en $Q$-normant $g$. Il vient alors:

$$(8a') \quad R_{X,M}PZQgg'QZ'PR_{X,M}Pf=\eta f$$

$$(8b') \quad R_{Y,N}QZ'Pff'PZQR_{Y,N}Qg=\eta g$$

Réciproquement: comme $g'QZ'PR_{X,M}Pf \in \mathbb{R}$, (8a') $\Rightarrow$ $R_{X,M}PZQg=\theta f$. De même, (8b') $\Rightarrow$ $R_{Y,N}QZ'Pf=\nu g$. Les coefficients de proportionnalité importent peu puisqu'on ne s'intéresse qu'aux directions. *In fine*, on peut donc caractériser ($f$,$g$) normés comme solution du système (8a',8b') avec $\eta$ maximale.

**c) Si l'on ne tient plus compte de la force structurelle des composantes:**

On utilise alors les métriques $M = (X'PX)^{-1}$ et $N = (Y'QY)^{-1}$. En effet, dans ce cas:

$$u'Mu=1 \Rightarrow \|XMu\|_P^2=1 \quad ; \quad v'Nv=1 \Rightarrow \|YNv\|_Q^2=1$$

Les composantes ayant alors toutes la même norme, les groupes $X$ et $Y$ n'interviennent qu'en tant que sous-espaces. Le programme est dans ce cas noté $\boldsymbol{P}(Z;\langle X\rangle,\langle Y\rangle)$. Ses solutions sont caractérisées par des équations particulières. En effet:

$$R_{X,M}P=X(X'PX)^{-1}X'P=\Pi_X \quad ; \quad R_{Y,N}Q=Y(Y'QY)^{-1}Y'Q=\Pi_Y$$

Donc:

$$(8a) \Leftrightarrow \Pi_X ZQ\Pi_Y Z'Pf=\eta f \quad \Leftrightarrow \quad \Pi_X ZQ\Pi_Y Z'P\Pi_X f=\eta f \quad (9a)$$

$$\text{car} \quad f\in\langle X\rangle \Rightarrow \Pi_X f=\eta f$$

La matrice $A=\Pi_X ZQ\Pi_Y Z'P\Pi_X$ est $P$-symétrique car, comme $P\Pi_X = (P\Pi_X)'$ et $Q\Pi_Y = (Q\Pi_Y)'$, on a: $PA = A'P$. Par conséquent, ses vecteurs propres $f$ sont $P$-orthogonaux. De même:

$$(8b) \Leftrightarrow \Pi_Y Z'P\Pi_X ZQg=\eta g \quad \Leftrightarrow \quad \Pi_Y Z'P\Pi_X ZQ\Pi_Y g=\eta g \quad (9b)$$

$$\text{car} \quad g\in\langle Y\rangle \Rightarrow \Pi_Y g=\eta g$$

La matrice $B=\Pi_Y Z'P\Pi_X ZQ\Pi_Y$ étant $Q$-symétrique, ses vecteurs propres $g$ sont $Q$-orthogonaux. Les composantes $f$ (resp. $g$) solutions des équations (9a,9b) forment ainsi un système orthogonal.

En outre, si l'on applique dans ce cas les résultats du §2.2.*b*, on peut caractériser le premier couple de solutions $(f,g)$ comme solution du système:

$$(9a') \quad \Pi_X ZQgg'QZ'P\Pi_X f=\eta f$$

$$(9b') \quad \Pi_Y Z'Pff'PZQ\Pi_Y g=\eta g$$

avec $\eta$ maximale.

**d) Quelques cas particuliers restituant des méthodes classiques**

***ACP***

• Si l'on prend: $X = I_n$ avec $M = P^{-1}$, $Y = I_p$ avec $N = Q^{-1}$ et $Z$ quelconque, les équations (8a) et (8b) ci-dessus deviennent:

$$ZQZ'Pf=\eta f \quad ; \quad Z'PZQg=\eta g$$

Elles correspondent donc à celles des ACP respectives de $Z,P,Q$ et $Z',Q,P$ (que l'on sait duales).

• Si l'on prend $Z$ quelconque, $X = Z$ avec $M = Q$, et $Y = Z'$ avec $N = P$, les équations (8a) et (8b) ci-dessus deviennent:

$$(ZQZ'P)^3 f=\eta f \quad ; \quad (Z'PZQ)^3 g=\eta g$$

Elles fournissent également les composantes principales respectives de $Z,P,Q$ et $Z',Q,P$.

*Méthodes à deux groupes*

• Si l'on prend $X$ et $Y$ de même dimension en ligne ($n=p$), $P = Q$ et $Z = P^{-1}$, le critère maximisé par le programme ***P*** devient, d'après (5): $v'NY'PXMu$ . Ce programme se réduit donc au programme suivant:

$$\boldsymbol{Q}(X,M;Y,N):\quad Max_{\substack{u'Mu=1\\ v'Nv=1}} \langle XMu | YNv \rangle_P$$

Il s'agit du programme fondamental des méthodes à deux groupes que sont l'analyse canonique, l'analyse des redondances maximales (alias ACPVI), l'analyse inter-batteries et l'analyse PLS (cf. par exemple [Bry 2001] ou [Tenenhaus 1998]). On montre, d'après (8a) et (8b) que les composantes $f = XMu$ et $g = YNv$ solutions de ce programme sont les solutions (correspondant à la valeur propre la plus élevée) des équations:

$$R_{X,M} P R_{Y,N} P f = \eta f \quad ; \quad R_{Y,N} P R_{X,M} P g = \eta g$$

*Analyses de tableaux de distances*

• Ici, les sujets, qui sont aussi les objets, sont équipondérés ($P = Q = I$), $Y = X$ et $M = N$. Prenons alors pour tableau $Z$ la matrice de produits scalaires reconstituée à partir d'une matrice $D$ donnant les distances deux à deux entre les sujets (on suppose la distance euclidienne). La matrice de produits scalaires étant symétrique, elle peut s'écrire, d'une infinité de manières: $Z = HH'$ (les colonnes de $H$ s'interprètent alors comme les coordonnées des sujets dans une base orthonormée quelconque; une possibilité est de prendre les composantes principales du nuage). On obtient alors les équations:

$$\text{(8a)} \Leftrightarrow \ (R_{X,M} HH')^2 f = \eta f \ \ ; \text{(8b)} \Leftrightarrow \ (R_{X,M} HH')^2 g = \eta g$$

Ceci montre que $f^t = g^t$ pour tout $t$. De plus, cette équation caractérise les composantes $f^t$ comme étant les solutions en $f$ du programme $Q(X,M;H,I)$. Il s'agit des composantes qui permettent de reconstituer, à partir des variables explicatives, la structure de distance correspondant à $Z$.

On note que la technique que nous proposons dans cet article permet de traiter le cas plus général où $Z$ code les produits scalaires entre deux ensembles distincts d'individus, et encore plus généralement le cas où à la place d'un produit scalaire, on utilise un indicateur non symétrique d'accord pour chaque couple d'individus.

## 2.3. Composantes de rangs 2 et ultérieurs

### a) Cas général

• On désire obtenir une décomposition approchée de $Z$ sous la forme d'une combinaison de vecteurs de $C_1$.

$$\hat{Z}_T = \sum_{s=1}^{T} \sum_{t=1}^{T} \omega_{st} U^{st} \quad \text{avec} \quad U^{st} = f^s g^{t\prime}$$

où $\{f^s\}_s$ et $\{g^t\}_t$ sont des systèmes orthonormés

• D'après le théorème 1, les $U^{st}$ sont des vecteurs normés, et d'après le théorème 2, ils sont deux à deux orthogonaux. En conséquence, $\{U^{st}\}_{s,t}$ est un système orthonormé. On

pourra donc écrire:

$$\|\hat{Z}_T\|_R^2 = \sum_{s=1}^{T} \sum_{t=1}^{T} \omega_{st}^2$$

***Calcul des composantes de rang 2***

- Soient $X_1 = \Pi_{\langle f^1 \rangle^\perp} X$ et $Y_1 = \Pi_{\langle g^1 \rangle^\perp} Y$ . Au rang 2, on cherche $f^2 \in \langle X_1 \rangle$ et $g^2 \in \langle Y_1 \rangle$ unitaires. Ces nouvelles composantes permettent de construire les quatre vecteurs orthonormés suivants:

$$U^{11} = f^1 g^1{}', U^{12} = f^1 g^2{}', U^{21} = f^2 g^1{}', U^{22} = f^2 g^2{}'$$

Soit $C_2$ l'ensemble suivant:

$$C_2 = \left\{ \sum_{s=1}^{2} \sum_{t=1}^{2} a_{st} f^s g^t{}' | f^2 \in \langle X_1 \rangle, g^2 \in \langle Y_1 \rangle \right\}$$

$C_2$ est un cône. Le théorème 3 assure alors que la recherche de l'ajustement OLS de $Z$ par un élément de $C_2$ équivaut à la recherche du vecteur unitaire de $C_2$ faisant avec $Z$ le cosinus maximum. Si l'on ne tenait pas compte de la force structurelle des composantes, il s'agirait donc de maximiser sur $f^2$ et $g^2$:

$$\cos^2(Z, \langle f^1 g^1{}', f^1 g^2{}', f^2 g^1{}', f^2 g^2{}' \rangle)$$

On procède de façon algorithmique. Considérons que $g^2$ ait été trouvé; il s'agit alors de déterminer $f^2$ maximisant:

$$\cos^2(Z, \langle f^1 g^1{}', f^1 g^2{}', f^2 g^1{}', f^2 g^2{}' \rangle)$$

Les vecteurs $U^{st} = f^s g^t{}'$ étant deux à deux orthogonaux, on a:

$$\cos^2(Z, \langle f^1 g^1{}', f^1 g^2{}', f^2 g^1{}', f^2 g^2{}' \rangle)$$

$$= \cos^2(Z, \langle f^1 g^1{}', f^1 g^2{}' \rangle) + \cos^2(Z, \langle f^2 g^1{}', f^2 g^2{}' \rangle)$$

Il s'agit donc de déterminer $f^2$ maximisant:

$$\cos^2(Z, \langle f^2 g^1{}', f^2 g^2{}' \rangle)$$

$$= \cos^2(Z, a_{21} f^2 g^1{}' + a_{22} f^2 g^2{}') = \cos^2(Z, f^2 (a_{21} g^1{}' + a_{22} g^2{}'))$$

On pose $G = (g^1, g^2)$ et $a' = (a_{21}, a_{22})$. On a donc: $a_{21} g^1 + a_{22} g^2 = Ga$. En prenant $a$ $I$-normé, on a: $Ga$ $Q$-normé.

D'autre part, il faut considérer la force structurelle de $f^2$. On pose donc $f^2 = X^1 M u_2$, et on prend $u_2$ normé. On doit donc résoudre:

$$\underset{\substack{u_2' M u_2 = 1 \\ a' a = 1}}{Max} \langle Z | X^1 M u_2 (Ga)' \rangle_R$$

De même, lorsque $f^2$ est déterminé, on calcule $g^2 = Y^1 N v_2$ en résolvant:

$$\underset{\substack{v_2' N v_2 = 1 \\ b' b = 1}}{Max} \langle Z | Fb (Y^1 N v_2)' \rangle_R \quad \text{où } F = (f^1, f^2)$$

La marche à suivre est donc simple:

- On détermine $f^1$ et $g^1$ en normant les solutions $f = XMu$ et $g = YNv$ du programme $\boldsymbol{P}(Z; (X,M),(Y,N))$. Puis, on calcule les résidus $X^1$ et $Y^1$:

$$X^1 = (I_n - f^1 f^{1\prime} P)X \quad ; \quad Y^1 = (I_p - g^1 g^{1\prime} Q)Y$$

- $f^2$ et $g^2$ sont déterminés itérativement à partir de valeurs initiales $f^2(0)$ et $g^2(0)$ en itérant jusqu'à convergence:

  (i) $f^2(k+1)$ = solution $f$ $P$-normée de: $\boldsymbol{P}(Z; (X^1,M),(G(k),I))$ où $G(k) = (g^1,g^2(k))$

  (ii) $g^2(k+1)$ = solution $g$ $Q$-normée de: $\boldsymbol{P}(Z; (F(k+1),I),(Y^1,N))$ où $F(k+1) = (f^1,f^2(k+1))$

***Algorithme FILM-A donnant les composantes jusqu'au rang T***

- Au rang $t$ , on calcule les groupes:

$$X^{t\text{-}1} = (I_n - f^{t\text{-}1} f^{t\text{-}1\prime} P)X^{t\text{-}2} \ ; \ Y^{t\text{-}1} = (I_p - g^{t\text{-}1} g^{t\text{-}1\prime} Q)Y^{t\text{-}2}$$

$$F^{t\text{-}1} = (f^1, \ldots, f^{t\text{-}1}) \ ; \ G^{t\text{-}1} = (g^1, \ldots, g^{t\text{-}1})$$

  Puis, on itère jusqu'à convergence:

  (i) $f^t(k+1)$ = solution $f$ $P$-normée de: $\boldsymbol{P}(Z; (X^{t\text{-}1},M),(G^t(k),I))$, où:

  $G^t(k) = (G^{t\text{-}1},g^t(k))$

  (ii) $g^2(k+1)$ = solution $g$ $Q$-normée de: $\boldsymbol{P}(Z; (F^t(k+1),I),(Y^{t\text{-}1},N))$, où:

  $F^t(k+1) = (F^{t\text{-}1}, f^t(k+1))$

- On régresse enfin $Z$ sur les vecteurs $U^{st} = f^s g^{t\prime}$. Ces vecteurs étant orthonormés, les coefficients de régression sont simplement:

$$\omega_{st} = \langle Z \mid U^{st} \rangle_R = \mathrm{tr}(Q g^t f^{s\prime} P Z) = f^{s\prime} P Z Q g^t$$

  On notera que l'on retrouve ici, étendue à deux systèmes de poids quelconques, la formule donnée par [Martens *et al.* 2005]:

$$\hat{\Omega} = (F'PF)^{-1} F'PZQG(G'QG)^{-1} = F'PZQG'$$

**b) Si l'on ne tient plus compte de la force structurelle des composantes**

• Lorsqu'on a choisi $M = (X'PX)^{-1}$ et $N = (Y'QY)^{-1}$, on a vu au §2.2.*c* que les composantes $f^t$ et $g^t$ solutions des conditions du 1^er^ ordre forment des systèmes orthogonaux. On montre ici que la base $\{U^{st}\}_{s,t}$ trouvée par l'algorithme FILM-A ci-dessus se réduit à:

$$\{U^t\}_t \quad \text{où} \quad U^t = f^t g^{t\prime} \ \ \forall t$$

Considérons la détermination des composantes de rang $t = 2$ (la généralisation aux rangs suivants n'est pas très difficile).

(a) Montrons d'abord que l'algorithme FILM-A fournit bien les composantes $(f^2, g^2)$ solutions des conditions du 1^er^ ordre:

Le pas (i) de l'itération $(k+1)$ de l'algorithme caractérise $f^2(k+1)$ comme solution $f$ normée de: $\boldsymbol{P}(Z; (X^1,M),(G(k),I))$ où $G(k) = (g^1,g^2(k))$. On a donc, en appliquant (9a):

$$\Pi_{X^1} ZQ(g^1 g^1{}' + g^2(k) g^2{}'(k)) QZ' P \Pi_{X^1} f^2(k+1) = \eta f^2(k+1) \quad (10)$$

$$\Leftrightarrow \quad \Pi_{X^1} ZQg^1 g^1{}' QZ' P \Pi_{X^1} f^2(k+1) + \Pi_{X^1} ZQg^2(k) g^2{}'(k) QZ' P \Pi_{X^1} f^2(k+1)$$

$$= \eta f^2(k+1)$$

Mais:

$$\Pi_{X^1} = \Pi_{X^1} \Pi_X = \Pi_X \Pi_{X^1} \quad \Rightarrow$$

$$\Pi_{X^1} ZQg^1 g^1{}' QZ' P \Pi_{X^1} f^2(k+1) = \Pi_{X^1} \Pi_X ZQg^1 g^1{}' QZ' P \Pi_X \Pi_{X^1} f^2(k+1)$$

Or, d'après (7a): $\Pi_X ZQg^1 \propto f^1$ . D'autre part, $\Pi_X$ étant $P$-symétrique, on a: $P \Pi_X = \Pi_X{}' P$ . Il s'ensuit:

$$\Pi_{X^1} \Pi_X ZQg^1 g^1{}' QZ' P \Pi_X \Pi_{X^1} f^2(k+1) \propto \Pi_{X^1} f^1 f^1{}' P \Pi_{X^1} f^2(k+1) = 0$$

$$\text{car} \quad \Pi_{X^1} f^1 = 0$$

Finalement:

$$(10) \Leftrightarrow \quad \Pi_{X^1} Z Q g^2(k) g^2(k)' Q Z' P \Pi_{X^1} f^2(k+1) = \eta f^2(k+1) \quad (11a)$$

Symétriquement, on établit que:

$$\Pi_{Y^1} Z' P f^2(k) f^2(k)' P Z Q \Pi_{Y^1} g^2(k+1) = \eta g^2(k+1) \quad (11b)$$

D'après (9a') et (9b') , (11a) et (11b) caractérisent $f^2(\infty)$ et $g^2(\infty)$ comme solutions du programme $\boldsymbol{P}(Z;\langle X^1\rangle;\langle Y^1\rangle)$. Or, ces dernières ne sont autres que les solutions de rang 2 des conditions du premier ordre du programme $\boldsymbol{P}(Z;\langle X\rangle;\langle Y\rangle)$ puisque $f^2 \in \langle X^1\rangle$ et $g^2 \in \langle Y^1\rangle$.

(b) Montrons à présent que les vecteurs $U^{st} = f^s g^t{}'$ où $s \neq t$ ont un coefficient nul dans la décomposition de $Z$.

Supposons d'abord $s>t$:

$$\langle Z | f^s g^t{}' \rangle_R = tr(QZ' P f^s g^t{}') = g^t{}' Q Z' P f^s = g^t{}' Q Z' P \Pi_{X^{t-1}} f^s$$

Or, $\Pi_{X^{t-1}}$ est $P$-symétrique, donc: $P \Pi_{X^{t-1}} = \Pi_{X^{t-1}}{}' P$ . En outre, on sait que (7a) entraîne ici: $\Pi_X^{t-1} ZQg^t \propto f^t$ . Il s'ensuit finalement:

$$\langle Z | f^s g^t{}' \rangle_R \propto f^t{}' P f^s = 0$$

Similairement, pour $s<t$, on a:

$$\langle Z | f^s g^t{}' \rangle_R = tr(QZ' P f^s g^t{}') = g^t{}' QZ' P f^s = g^t{}' Q \Pi_{Y^{s-1}} Z' P f^s \propto g^t{}' Q g^s = 0$$

• En conclusion, comme dans ce cas particulier il n'y a pas de termes d'interactions "croisées", la décomposition de $Z$ est particulièrement parcimonieuse. Les composantes obtenues dans ce cas très particulier sont celles de l'analyse RLQ.

### 2.4. Décomposition finale et indicateurs utiles

• Ayant calculé $T^2$ vecteurs $U^{st}$ orthonormés, on a:

$$\hat{Z}_T = \sum_{s=1}^{T} \sum_{t=1}^{T} \omega_{st} U^{st} = \Pi_{\langle \{U^{st}\}_{s,t} \rangle} Z \quad \text{et} \quad \hat{E}_T = Z - \hat{Z}_T = \Pi_{\langle \{U^{st}\}_{s,t} \rangle^{\perp}} Z$$

• On a, en appliquant Pythagore:

$$\|\hat{Z}_T\|_R^2 = \sum_{s,t} \omega_{st}^2 \quad \text{et} \quad \|Z\|_R^2 = \|\hat{Z}_T\|_R^2 + \|\hat{E}_T\|_R^2$$

• Les contributions des termes de la décomposition à la norme carrée de $Z$ seront mesurées en proportion de celle-ci. On calculera ainsi:

$$\frac{\|\hat{Z}_T\|_R^2}{\|Z\|_R^2} \quad ; \quad \forall s,t: \frac{\omega_{st}^2}{\|Z\|_R^2} \quad ; \quad \frac{\|\hat{E}_T\|_R^2}{\|Z\|_R^2}$$

• On peut également, *via* Pythagore, décomposer $X$ (et $\|X\|^2$) sur ses composantes $f^s$, et $Y$ (et $\|Y\|^2$) sur ses composantes $g^t$.

Les parts de variance expliquée par les composantes sont dans tous les cas additives, par orthogonalité. La part de variance de $X$ expliquée par $f^s$, par exemple, est:

$$\frac{f^{s\prime} PXX'Pf^s}{tr(X'PX)} = \frac{f^{s\prime} PXX'Pf^s}{J} \quad \text{si les } x^j \text{ sont normées.}$$

# 3. Application à l'analyse des interactions

### 3.1. Centrages et conséquences

#### a) Centrage des composantes et centrages de *Z* en ligne et colonne

Les variables explicatives $x^j$ et $y^k$ étant en pratique hétérogènes, elles doivent être standardisées préalablement à l'analyse. Typiquement, elles seront centrées et réduites, afin d'être analysées en termes de corrélation linéaire. Le centrage des variables impose une contrainte forte à la décomposition recherchée de $Z$, ce qui rend le modèle A irréaliste si $Z$ n'est pas centré en ligne et en colonne, donc en général.

• Prenons le cas élémentaire d'un tableau de rang 1. Le centrage de $f$ et celui de $g$ implique que $U = fg'$ est un tableau centré en ligne et en colonne:

$$UQe_p = f(g'Qe_p) = 0 \quad ; \quad e_n'PU = (e_n'Pf)g' = 0$$

• Réciproquement: si $U = fg'$ avec $f$ et $g$ non centrés, et soient $f^* = f$ centré, $g^* = g$ centré, et $U^*=U$ centré en ligne et en colonne; on a: $U^* = f^*g^{*\prime}$.

Notons:

$$\forall i,\ \bar{u}_i = \sum_j q_j u_i^j \quad ; \quad \forall j,\ \bar{u}^j = \sum_i p_i u_i^j \quad ; \quad \bar{\bar{u}} = \sum_{i,j} p_i q_j u_i^j = \sum_j q_j \bar{u}^j = \sum_i p_i \bar{u}_i$$

On a:

$$U = fg' \Leftrightarrow \forall i, j : u_i^j = f_i g_j = (f_i^* + \bar{f})(g_j^* + \bar{g})$$

$$\Leftrightarrow \quad \forall i, j: \quad u_i^j = f_i^* g_j^* + \bar{g} f_i^* + \bar{f} g_j^* + \bar{f} \bar{g}$$

$$\Leftrightarrow \quad \forall i, j: \quad u_i^j = f_i^* g_j^* + \bar{g}(f_i - \bar{f}) + \bar{f}(g_j - \bar{g}) + \bar{f} \bar{g}$$

$$\Leftrightarrow \quad \forall i, j: \quad u_i^j = f_i^* g_j^* + \bar{g} f_i + \bar{f} g_j - \bar{f} \bar{g}$$

Or:

$$\forall i: \quad \bar{u}_i = \sum_j q_j f_i g_j = \bar{g} f_i \quad ; \quad \forall j: \quad \bar{u}^j = \sum_i p_i f_i g_j = \bar{f} g_j$$

$$\bar{\bar{u}} = \sum_{i,j} p_i q_j f_i g_j = \bar{f} \bar{g}$$

Donc:

$$\forall i, j: \quad u_i^j - \bar{u}_i - \bar{u}^j + \bar{\bar{u}} = f_i^* g_j^*$$

**b) Ajustement du modèle A et centrage**

Théorème 4:

L'application de FILM-A à $Z$ sur des tableaux $X$ et $Y$ de variables centrées équivaut à son application sur le tableau $Z^*$ obtenu en centrant $Z$ en ligne et en colonne.

Preuve:

Le centrage de $Z$ en ligne et en colonne se fait comme suit:

$$\forall i, j: \quad z_i^{j*} = z_i^j - \bar{z}_i - \bar{z}^j + \bar{\bar{z}} \quad \Leftrightarrow \quad Z^* = Z - ZQe_p e_p' - e_n e_n' PZ + \bar{\bar{z}} e_n e_p'$$

Or:

$$\langle XMuv' N' Y' | Z^* \rangle_R = tr(QYNvu' MX' PZ^*) = u' MX' PZ^* QYNv$$

De plus:

$$X' PZ^* QY = X' PZQY - X' PZQe_p e_p' QY - X' Pe_n e_n' PZQY + \bar{\bar{z}} X' Pe_n e_p' QY$$

Et comme le centrage de $X$ et $Y$ équivaut respectivement à la nullité de $X'Pe_n$ et $Y'Qe_p$:

$$X' PZ^* QY = X' PZQY$$

Donc:

$$\langle XMuv' N' Y' | Z^* \rangle_R = \langle XMuv' N' Y' | Z \rangle_R$$

## *3.2. Analyse d'un tableau avec effets propres et interactions (modèles B1 et B2)*

Nous voyons ci-dessous qu'un tableau $Z$ quelconque (non centré en ligne et colonne) apporte une triple information:

|- Une information marginale de disparité entre sujets, "indépendante" de l'objet;

- Une information marginale de disparité entre objets, "indépendante" du sujet;

- Une information d'interaction "pure" entre sujets et objets, correspondant à un tableau centré en ligne et colonne.

Dans la mesure du possible, il nous faut chercher des composantes permettant d'exprimer au mieux les informations des trois types. Nous verrons que deux stratégies sont envisageables.

**a) Décomposition d'un tableau *Z* quelconque**

• Considérons deux vecteurs $f \in \mathbb{R}^n$ et $g \in \mathbb{R}^p$. On note $f^*$ et $g^*$ les vecteurs $f$ et $g$ centrés. On a:

$$fg' = (f^* + \bar{f} e_n)(g^* + \bar{g} e_p)' = f^* g^{*\prime} + \bar{f} \bar{g} e_n e_p' + \bar{g} f^* e_p' + \bar{f} e_n g^{*\prime}$$

Or, $f^*$ et $g^*$ étant centrés, les quatre vecteurs de cette somme sont deux à deux orthogonaux. Démontrons, à titre d'exemple, trois des six orthogonalités en question, les trois dernières se montrant de façon analogue:

$$\langle f^* g^{*\prime} | f^* e_p' \rangle_R = tr(Q g^* f^{*\prime} P f^* e_p') = e_p' Q g^* f^{*\prime} P f^* = 0 \; f^{*\prime} P f^* = 0$$

$$\langle f^* g^{*\prime} | e_n e_p' \rangle_R = tr(Q g^* f^{*\prime} P e_n e_p') = e_p' Q g^* f^{*\prime} P e_n = 0$$

$$\langle f^* e_p' | e_n e_p' \rangle_R = tr(Q e_p f^{*\prime} P e_n e_p') = e_p' Q e_p f^{*\prime} P e_n = 0$$

• Considérons à présent un tableau $Z$ pouvant se décomposer la forme suivante:

$$Z = \sum_{t=1}^{T} f^t g^{t\prime} \quad \text{où} \quad f^t \in \mathbb{R}^n \text{ et } g^t \in \mathbb{R}^p \text{ quelconques}$$

On obtient, en faisant apparaître les $f^t$ et $g^t$ centrées:

$$Z = \sum_{t=1}^{T} \left( f^{t*} g^{t*\prime} + \bar{f}^t \bar{g}^t e_n e_p' + \bar{g}^t f^{t*} e_p' + \bar{f}^t e_n g^{t*\prime} \right)$$

$$Z = \left( \sum_{t=1}^{T} \bar{f}^t \bar{g}^t \right) e_n e_p' + \left( \sum_{t=1}^{T} \bar{g}^t f^{t*} \right) e_p' + e_n \left( \sum_{t=1}^{T} \bar{f}^t g^{t*} \right)' + \sum_{t=1}^{T} f^{t*} g^{t*}$$

Encore une fois, on obtient une somme de quatre vecteurs orthogonaux deux à deux.

• Passons enfin à un tableau $Z = ((z_i^j))_{i,j}$ quelconque. Il est décomposable de façon unique en:

$$Z = Z^0 + Z^i + Z^m + Z^* \quad \text{où:}$$

$Z^0 = \alpha e_n e_p'$ ; $Z^*$ centré en ligne et colonne ;

$Z^i = f e_p'$ où $f \in \mathbb{R}^n$ $P$-centré ($e_n' P f = 0$) ;

$Z^m = e_n g'$ où $g \in \mathbb{R}^p$ $Q$-centré ($e_p' Q g = 0$)

Preuve:

On a: $e_n' P e_n = 1$ et $e_p' Q e_p = 1$. Le centrage de $Z^*$ en ligne et colonne implique:

$$e_n'PZ^*Qe_p = 0 \quad \Leftrightarrow \quad e_n'P(Z - \alpha e_n e_p' - f e_p' - e_n g')Qe_p = 0 \quad \Leftrightarrow \quad \alpha = \bar{\bar{z}}$$

D'autre part:

$$e_n' PZ^* = 0 \quad \Leftrightarrow \quad e_n' PZ - \bar{\bar{z}} e_p' - g' = 0$$

$$\Leftrightarrow \quad g = Z' Pe_n - \bar{\bar{z}} e_p \quad \Leftrightarrow \quad \forall m, g_m = \bar{z}^m - \bar{\bar{z}}$$

Et de même:

$$Z^* Q e_p = 0 \quad \Leftrightarrow \quad \forall i, f_i = \bar{z}_i - \bar{\bar{z}}$$

Ce qui entraîne enfin:

$$z_i^{m*} = z_i^m - (\bar{z}_i - \bar{\bar{z}}) - (\bar{z}^m - \bar{\bar{z}}) - \bar{\bar{z}} = z_i^m - \bar{z}_i - \bar{z}^m + \bar{\bar{z}}$$

## b) Stratégies d'analyse

La décomposition donnée en (a) permet d'envisager la modélisation factorielle séparée des tableaux $Z^i$ , $Z^m$ et $Z^*$ (modélisation notée B1) ou au contraire leur modélisation simultanée (modélisation notée B2).

N.B.: les variables de $X$ et $Y$ sont prises centrées.

### *Modélisation séparée de $Z^i$ , $Z^m$ et $Z^*$ (FILM-B1)*

On utilise un programme de régression OLS1 ou PLS1 pour modéliser le $f$ de $Z^i$ en fonction de composantes $f_1^{t*}$ $P$-orthonormées de <$X$> (resp. $g$ de $Z^m$ en fonction de composantes $g_1^{t*}$ $Q$-orthonormées de <$Y$>) et l'algorithme FILM-A (cf. §2.3) pour décomposer $Z^*$ en $\sum_{s,t} \omega_{st} f^{s*} g^{t*}{}'$ (utilisant aussi des composantes orthonormées). Cette façon de procéder n'impose aucune contrainte d'identité des composantes $f_1^{s*}$ (resp. $g_1^{t*}$) et $f^{s*}$ (resp. $g^{t*}$).

On aura, après l'analyse séparée de chaque vecteur:

$$Z^i = \left( \sum_{t=1}^{T} \alpha_t f_1^t * \right) e_p' + W_1 \quad ; \quad Z^m = e_n \left( \sum_{t=1}^{T} \beta_t g_1^t * \right)' + W_2 \quad ;$$
$$Z^* = \sum_{s,t} \omega_{st} f^s * g^t *' + W_3$$

Le résidu $W$ est dans chaque cas orthogonal à la partie expliquée. Les vecteurs $W_1, W_2, W_3$ sont aussi orthogonaux entre eux. Ces orthogonalités permettent de décomposer additivement la variance de $Z$:

$$\|Z - \bar{Z} e_n e_p'\|_R^2 = \|Z^i\|_R^2 + \|Z^m\|_R^2 + \|Z^*\|_R^2 \quad =$$

$$\left\| \left( \sum_{t=1}^{T} \alpha_t f_1^t * \right) e_p' \right\|_R^2 + \|W_1\|_R^2 + \left\| e_n \left( \sum_{t=1}^{T} \beta_t g_1^t * \right)' \right\|_R^2 + \|W_2\|_R^2 + \left\| \sum_{s,t} \omega_{st} f^s * g^t *' \right\|_R^2 + \|W_3\|_R^2$$

La décomposition va même plus loin:

$$\left\| \left( \sum_{t=1}^{T} \alpha_t f_1^t * \right) e_p' \right\|_R^2 = e_p' Qe_p \left( \sum_{t=1}^{T} \alpha_t f_1^t * \right)' P \left( \sum_{t=1}^{T} \alpha_t f_1^t * \right) = \sum_{t=1}^{T} \alpha_t^2 (f_1^t *' P f_1^t *) = \sum_{t=1}^{T} \alpha_t^2$$

De même:

$$\left\| e_n \left( \sum_{t=1}^{T} \beta_t g_1^t * \right)' \right\|_R^2 = \sum_{t=1}^{T} \beta_t^2$$

Enfin:

$$\left\| \sum_{s,t} \omega_{st} f^s * g^t *' \right\|_R^2 = \sum_{s,t} \omega_{st}^2$$

***Modélisation simultanée de $Z^i$ , $Z^m$* et *$Z^*$ (FILM-B2)***

La seconde stratégie envisageable consiste à traiter *ensemble* les tableaux de la décomposition de sorte à imposer l'identité des composantes $f$ (resp. $g$) trouvées pour décomposer $Z^i$ (resp. $Z^m$) et $Z^*$.

- **Composantes de rang 1:**

Nous allons tout d'abord envisager la stratégie dans le cas où les composantes $f$ et $g$ n'ont pas de contrainte de force structurelle dans les groupes $X$ et $Y$. Dans un deuxième temps, nous introduirons de telles contraintes.

Supposons que l'on cherche à ajuster $\beta f e_p' + \gamma e_n g' + \delta f g'$ à: $Z - \bar{\bar{z}} e_n e_p'$ , $f$ et $g$ étant deux vecteurs unitaires centrés de $\langle X \rangle \subset \mathrm{IR}^n$ et $\langle Y \rangle \subset \mathrm{IR}^p$ respectivement. Nous procéderons de façon itérative, en maximisant l'ajustement alternativement sur $f$ et sur $g$:

- A $f$ et $\beta$ fixés, on cherche à ajuster au mieux $\gamma e_n g' + \delta f g'$ à : $Z - \bar{\bar{z}} e_n e_p' - \beta f e_p'$ . On cherche donc à résoudre:

$$Max_{\gamma,\delta,g/g'g=1} \cos_R \left( Z - \bar{\bar{z}} e_n e_p' - \beta f e_p' \; ; \; (\gamma e_n + \delta f) g' \right)$$

Notons $d = \begin{pmatrix} \gamma \\ \delta \end{pmatrix}$ et $H = (e_n^* , f)$ avec $e_n^* = e_n$ $P$-normé. On peut toujours imposer à $d$ d'être $I$-normé, ce qui permet à $h=Hd$ d'être $P$-normé. Le programme s'écrit alors:

$$\underset{\substack{h \in \langle H \rangle, h'Ph=1 \\ g \in \langle Y^* \rangle, g'Qg=1}}{Max} \cos_R \left( Z - \bar{\bar{z}} e_n e_p' - \beta f e_p' \; ; \; h g' \right)$$

Si l'on désire à présent imposer à $g$ une contrainte de force structurelle dans <$Y$>, le programme devient:

$$\underset{\substack{d, d'd=1 \\ v, v'Nv=1}}{Max} \left\langle Z - \bar{\bar{z}} e_n e_p' - \beta f e_p' \mid Hd(YNv)' \right\rangle_R$$

Ce n'est rien d'autre que le programme:

$$\boldsymbol{P}(Z - \bar{\bar{z}} e_n e_p' - \beta f e_p' \; ; \; (H, I), (Y, N))$$

- De même, à $g$ et $\gamma$ fixés, on calculera $f$ comme la solution $P$-normée du programme:

$$\boldsymbol{P}(Z - \bar{\bar{z}} e_n e_p' - \gamma e_n g' \; ; \; (X, M), (K, I)) \quad \text{où } K = (e_p^*, g)$$

- **Composantes de rangs 2 et ultérieurs**

On régresse aisément $Z - \bar{\bar{z}} e_n e_p'$ sur $f^1 e_p', e_n g^{1\prime}$ et $f^1 g^{1\prime}$ (ces deux derniers vecteurs étant orthogonaux), ce qui fournit un résidu $Z^1$.

On va chercher dans $X^1$ une composante $f^2$ orthogonale à $f^1$ et dans $Y^1$ une composante $g^2$ orthogonale à $g^1$ pour poursuivre la décomposition de $Z$.

Le résidu $Z^1$, orthogonal à $f^1 e_p', e_n g^1{}'$ et $f^1 g^1{}'$ doit être décomposé selon les vecteurs orthogonaux de $\mathbf{R}^{np}$: $f^2 e_p', e_n g^2{}', f^2 g^2{}', f^1 g^2{}', f^2 g^1{}'$ . Ce sont les vecteurs $f^1 g^2{}', f^2 g^1{}'$ qui compliquent un peu la tâche. On procédera comme pour les composantes de rang 2 dans l'algorithme FILM-A (§2.3):

Pour ajuster à $Z^1$ un vecteur:

$$\alpha f^2 e_p' + \beta e_n g^2{}' + \gamma_{22} f^2 g^2{}' + \gamma_{12} f^1 g^2{}' + \gamma_{21} f^2 g^1{}'$$

- A $f^2$, $\alpha$ et $\gamma_{21}$ fixés, on ajustera à $Z^1 - \alpha f^2 e_p' - \gamma_{21} f^2 g^1{}'$ un vecteur $(\beta e_n + \gamma_{22} f^2 + \gamma_{12} f^1) g^2{}'$ . En posant $H = (e_n^*, f^1, f^2)$ (3 vecteurs orthonormés) et $d = \begin{pmatrix} \beta \\ \gamma_{12} \\ \gamma_{22} \end{pmatrix}$ que l'on $I$-norme, on calculera $g^2$ en résolvant le programme:

$$\boldsymbol{P}(Z^1 - \alpha f^2 e_p' - \gamma_{21} f^2 g^1{}' \ ; \ (H, I), (Y^1, N))$$

Cette résolution fournit $g^2 = Y^1 N v$, $\beta$, $\gamma_{12}$ et $\gamma_{22}$.

- De même, à $g^2$, $\beta$ et $\gamma_{12}$ fixés, on calculera $f^2$ en résolvant le programme:

$$\boldsymbol{P}(Z^1 - \beta e_n g^2{}' - \gamma_{12} f^1 g^2{}' \ ; \ (X^1, M), (K, I)) \quad \text{où } K = (e_p^*, g^1, g^2)$$

... ce qui donne $f^2 = X^1 M u$, $\alpha$, $\gamma_{21}$ et $\gamma_{22}$.

***En pratique***

On emploiera FILM-B1 *ab initio*, et si l'on constate une certaine convergence entre des composantes explicatives des effets propres et celles des interactions, on adoptera FILM-B2 pour estimer un modèle plus parcimonieux.

## *3.3. Application à la modélisation de la dépendance entre deux variables nominales*

Nous nous trouvons ici exactement dans la situation traitée par la méthode RLQ, le tableau d'interactions étant un tableau de contingence. Au contraire de RLQ, FILM modélise $Z$ en prenant en compte les interactions croisées entre composantes.

**a) Données et notations spécifiques:**

Soit $A = (f_{im})_{\substack{i=1,\dots n \\ m=1,\dots p}}$ le tableau de fréquences normalisé codant la distribution d'un couple de deux caractères qualitatifs $C1$ et $C2$. On dispose par ailleurs d'un tableau $X$ (resp. $Y$) de variables numériques décrivant les modalités de $C1$ (resp. $C2$).

A la modalité $i$ (resp. $m$) du caractère $C1$ (resp. $C2$) on associe un poids égal à sa fréquence marginale $f_{i.}$ (resp. $f_{.m}$), où:

$$f_{i.} = \sum_{m=1}^{p} f_{im} \quad ; \quad f_{.m} = \sum_{i=1}^{n} f_{im}$$

On note respectivement les matrices de poids: $P = \mathrm{diag}(f_{i.})_{i=1,\dots n}$, $Q = \mathrm{diag}(f_{.m})_{m=1,\dots p}$.

Les variables $x^j$ (resp. $y^k$) sont centrées-réduites relativement au système de poids $P$ (resp. $Q$).

**b) Codage matriciel de la liaison entre les caractères**

• Considérons le tableau $\Phi$ suivant:

$$\Phi=(\phi_{im})_{i,m} \quad \text{avec} \quad \phi_{im}=\frac{f_{im}}{f_{i.}f_{.m}}-1$$

• Premièrement, on a:

$$\|\Phi\|_R^2=tr(Q\Phi' P\Phi)=\sum_{i,m}\frac{(f_{im}-f_{i.}f_{.m})^2}{f_{i.}f_{.m}}$$

On reconnaît là le coefficient $\phi^2$ mesurant l'intensité de la liaison entre les deux caractères.

• Deuxièmement, l'ACB du tableau $A$ fournit la formule de reconstitution suivante:

$$f_{im}=f_{i.}f_{.m}\left(1+\sum_k\sqrt{\lambda_k}f_i^k g_m^k\right)$$

où l'on constate que le tableau $\Phi$ a fait l'objet d'un codage factoriel : $\phi_{im}=\sum_k\sqrt{\lambda_k}f_i^k g_m^k$ , où $\lambda_k$ est la $k$-ième valeur propre de l'ACB, $f^k$ et $g^k$ étant les composantes directe et duale normées correspondantes.

La méthode FILM va permettre de modéliser $\Phi$ à partir des variables de $X$ et $Y$.

**c) Application de FILM**

• On constate que $\Phi$ est centré en colonne pour les poids $f_{i.}$ et en ligne pour les poids $f_{.j}$. Pour le centrage en colonne, par exemple:

$$\sum_i f_{i.}\phi_{ij}=\frac{\sum_i f_{ij}}{f_{.j}}-\sum_i f_{i.}=\frac{f_{.j}}{f_{.j}}-1=1-1=0$$

Il en va de même pour le centrage en ligne, par symétrie.

On peut donc appliquer à $\Phi$ modèle et algorithme FILM-A:

$$\phi_{im}=\sum_k\omega_k f_i^k g_m^k$$

... où les $f^k$ sont cette fois des composantes $P$-centrées dans <$X$> et les $g^k$ des composantes $Q$-centrées dans <$Y$>.

Les composantes de rang 1 de RLQ et FILM coïncident, mais les composantes fournies par les deux méthodes diffèrent en général à partir du rang 2, à cause des interactions croisées.

# 4. Exemples

## *4.1. Simulation*

Nous avons réalisé 100 fois l'expérience suivante:

1) Générer $X$ (respectivement $Y$) selon le patron de *faisceaux orthogonaux* suivant:

- 3 faisceaux “explicatifs” F1, F2 et F3 (resp. G1, G2 et G3) de respectivement 3, 2 et 1 variables.
- Un faisceau “parasite” F4 (resp. G4) de 4 variables très fortement corrélées.
- Des vecteurs de bruit indépendants constituant une matrice $E$.

Les variables du faisceau F$k$ (resp. G$k$) sont engendrées en ajoutant autant de bruits de variance petite devant 1 à une même variable normée $f_k$ (resp. $g_k$).

2) Générer le tableau $Z$ comme suit:

$$Z = Z^* + E \quad \text{où:} \quad Z^* = \omega_1 f^1 g^{1\prime} + \omega_2 f^2 g^{2\prime} + \omega_3 f^3 g^{3\prime}$$

Ici: $$Z^* = .49 f^1 g^{1\prime} + .69 f^2 g^{2\prime} + .53 f^3 g^{3\prime}$$

3) On analyse $Z$ par FILM, d'abord sans tenir compte de la force structurelle des composantes, puis en en tenant compte. On note à chaque estimation le $R^2$, ainsi que, pour tout $k$: $\rho(f^k, \hat{f}^k)$ , $\rho(g^k, \hat{g}^k)$ .

De la première à la dernière expérience, nous avons fait croître la variance de $E$ de 0 à V($Z^*$). Le seuil de convergence des composantes est fixé à $10^{-9}$.

**a) Analyse sans tenir compte de la force structurelle:**

Le nombre d'itérations pour convergence a toujours été au plus égal à 7.

- L'évolution du $R^2$ selon l'amplitude du bruit est donnée par la figure 3.

*Figure 3: Evolution du $R^2$ selon l'amplitude du bruit*

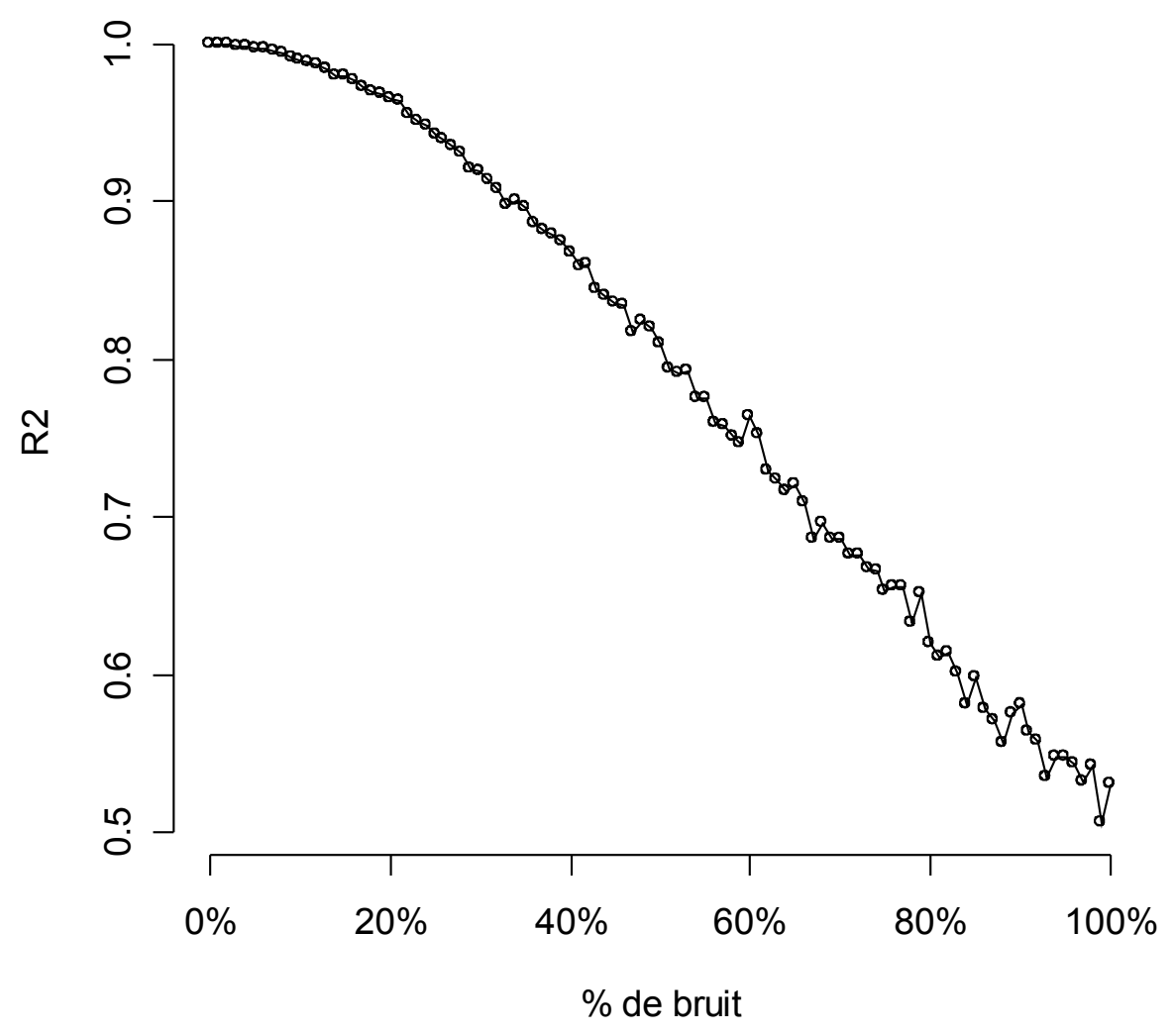

- L'évolution, selon l'amplitude du bruit, des corrélations entre facteurs originels et estimés est donnée par la figure 4.

*Figure 4: Evolution, selon l'amplitude du bruit, des corrélations* $\rho(f^k, \hat{f}^l)$

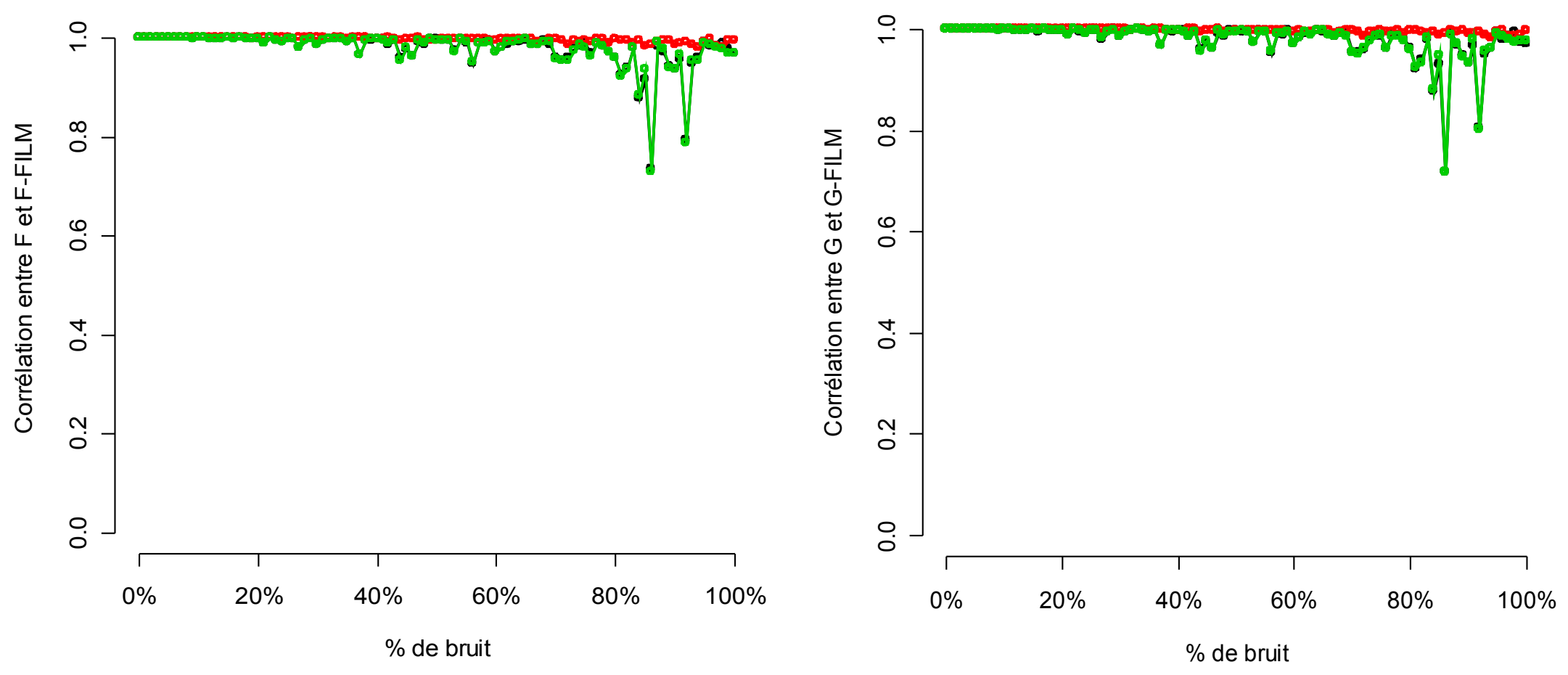

Noir : $\rho(f^1, \hat{f}^3)$ ; Rouge : $\rho(f^2, \hat{f}^1)$ ; Vert : $\rho(f^3, \hat{f}^2)$

La composante $f^2 g^{2'}$ est trouvée par FILM au rang 1, la composante $f^3 g^{3'}$ au rang 2, et la composante $f^1 g^{1'}$ au rang 3. Elles sont donc trouvées dans l'ordre du $\omega$ décroissant (donc de l'ajustement décroissant).

On constate que les corrélations restent élevées même lorsque le bruit $E$ a la même variance que le signal $Z^*$.

La matrice des coefficients estimés $\omega_{st}$ des vecteurs $U^{st}$ correspondant aux deux situations extrêmes en matière de bruit est:

| Bruit à 0%: | Bruit à 100%: |
|---|---|
| **0.691** 0.000 0.000<br>0.000 **0.531** 0.000<br>0.000 0.000 **0.491** | **0.508** 0.000 0.000<br>0.000 **0.379** 0.000<br>0.000 0.000 **0.343** |

**b) Analyse tenant compte de la force structurelle:**

- L'évolution du $R^2$ selon l'amplitude du bruit est donnée par la figure 5.

*Figure 5: Evolution du $R^2$ selon l'amplitude du bruit*

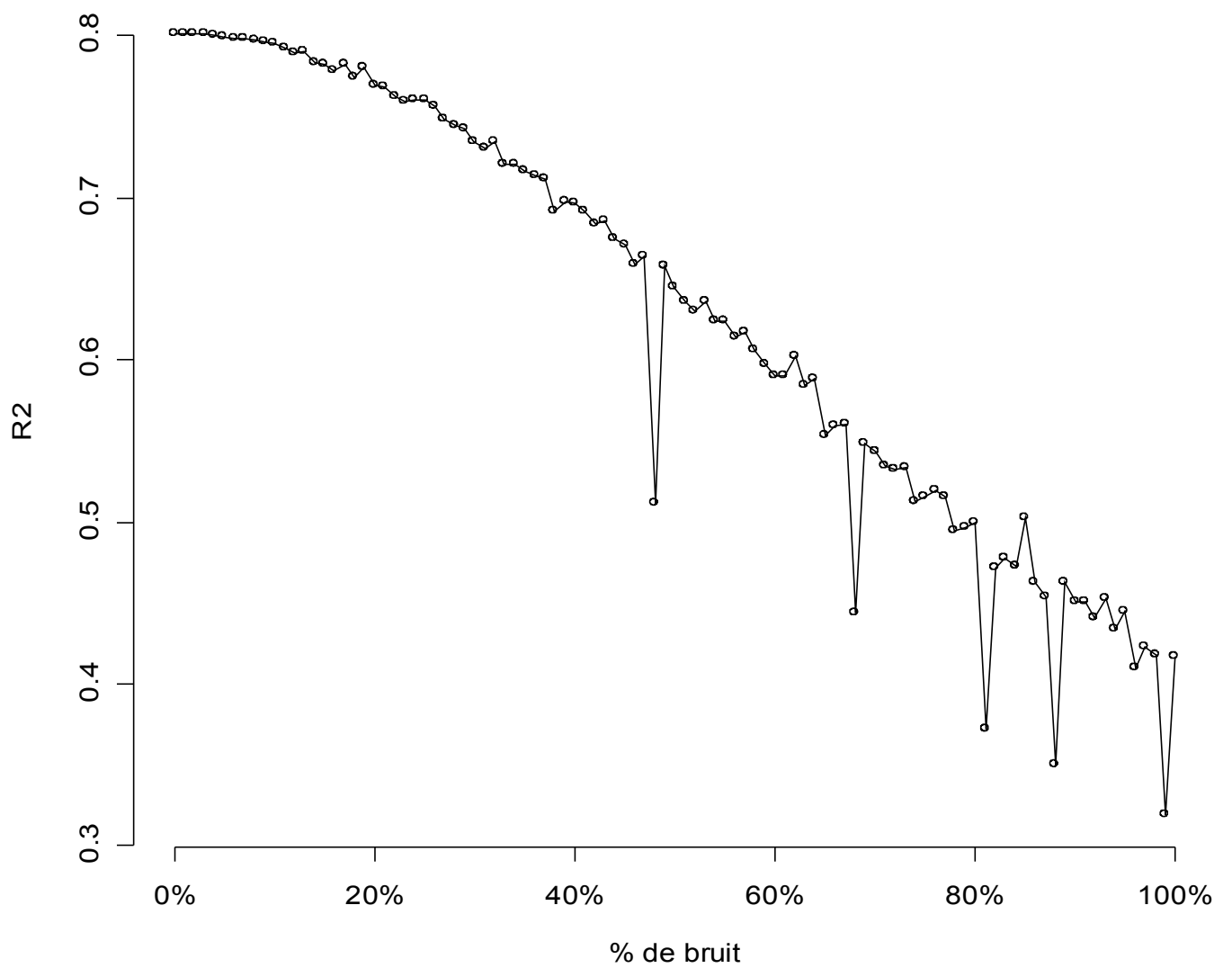

- L'évolution, selon l'amplitude du bruit, des corrélations entre facteurs originels et estimés est donnée par la figure 6.

*Figure 6: Evolution, selon l'amplitude du bruit, des corrélations* $\rho(f^k, \hat{f}^l)$

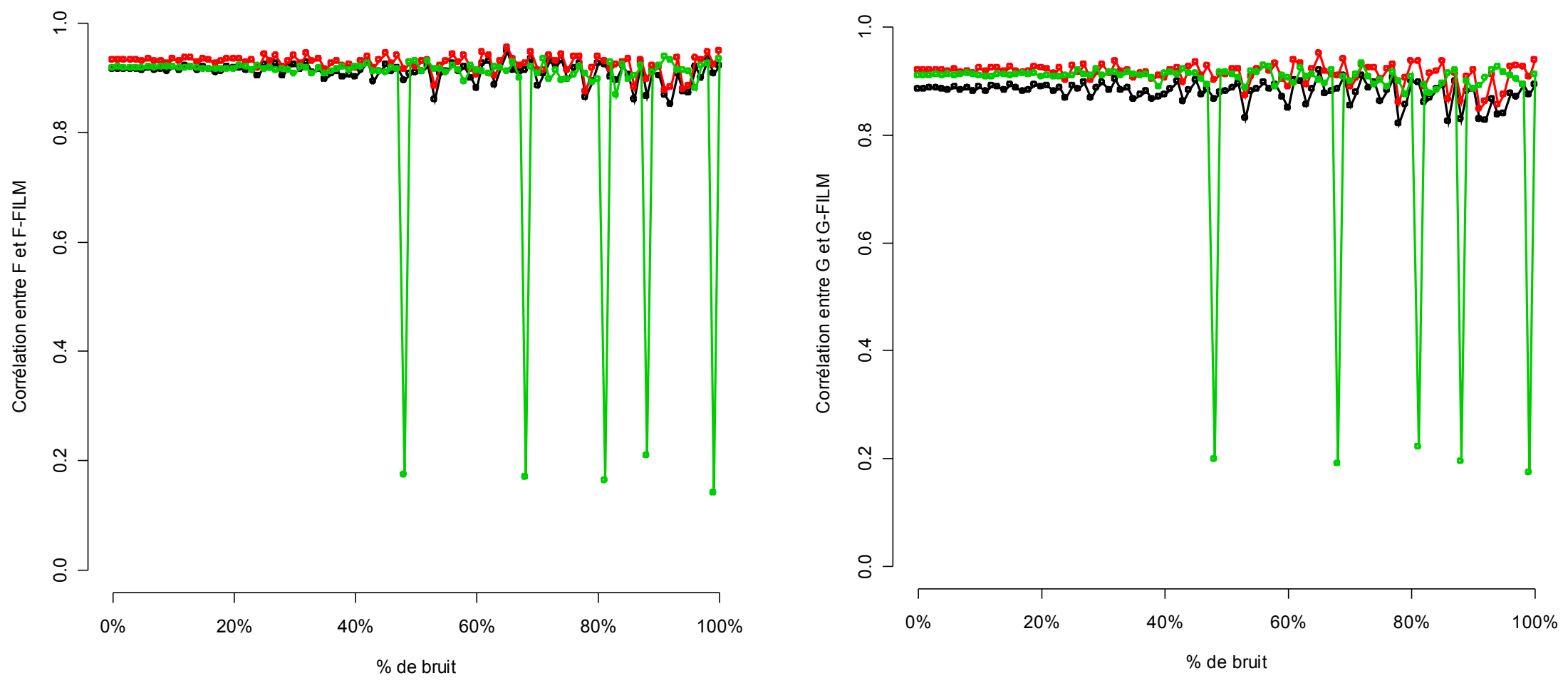

Noir : $\rho(f^{1,} \hat{f}^2)$ ; Rouge : $\rho(f^{2,} \hat{f}^1)$ ; Vert : $\rho(f^{3,} \hat{f}^3)$

La composante $f^2g^{2'}$ est trouvée par FILM au rang 1, la composante $f^1g^{1'}$ au rang 2, et la composante $f^3g^{3'}$ au rang 3. Elles ne sont donc plus trouvées dans l'ordre du $\omega$ décroissant - donc de l'ajustement décroissant - car cette fois, la force structurelle de la composante intervient. La composante la plus faible ($f^3g^{3'}$) est reléguée au troisième rang, bien qu'elle permette un ajustement légèrement meilleur que $f^1g^{1'}$, trouvée au rang 2.

On constate que les corrélations restent élevées même lorsque le bruit $E$ a la même variance que le signal $Z^*$, sauf dans les 5 cas caractérisés par un effondrement brutal du $R^2$. Il est facile d'analyser ceux-ci dès que l'on apprend que le vecteur $U^{33}$ est dans ces 5 cas très proche du faisceau 4, structurellement fort mais ne jouant pas de rôle dans $Z^*$: le bruit $E$, trop fort, a dans ce cas créé un "mirage explicatif partiel" de $Z$ par le faisceau 4, suffisant pour que cette structure, par sa force, soit dépistée prioritairement au faisceau 3.

La matrice des coefficients estimés $\omega_{st}$ des vecteurs $U^{st}$ correspondant aux deux situations extrêmes en matière de bruit est:

| Bruit à 0%: | Bruit à 100%: |
|---|---|
| **0.618** 0.069 -0.012 | **0.443** 0.034 -0.007 |
| 0.025 **0.449** -0.004 | 0.006 **0.322** -0.022 |
| -0.010 0.009 **0.460** | 0.007 -0.004 0.332 |

## *4.2. Micro-application sur données réelles: analyse des goûts musicaux*

N.B. Cette application n'a pour but que d'illustrer le fonctionnement de la méthode. Les données n'en ont pas été recueillies avec la rigueur qu'exigerait une véritable étude socio-musicologique, même élémentaire.

Dans cette application, nous décrivons:

- 18 individus (sujets), par des variables démographiques (*sexe, âge, statut matrimonial, nombre d'enfants*), économiques (*activité, revenu*) et socio-culturelles (*niveau d'études, pratique de la musique, pratique de la danse, écoute de musique sur CD, écoute de musique à la radio*). Les variables qualitatives sont codées par leurs indicatrices. Le nombre des variables numériques $x^j$ obtenues au total est de 21.
- 20 genres musicaux (objets), par les scores $y^k$ que leur ont attribué, sur une échelle ordinale comportant 5 degrés, un jury de mélomanes relativement à 17 critères: *complexite globale, complexité de la forme, complexité du rythme, complexité harmonique, complexité mélodique, complexité des textes* (0 en cas d'absence), *récursivité, répétitivité, importance du rythme, importance de l'harmonie, importance de la mélodie, richesse instrumentale, dansabilité, degré de contrainte stylistique, durée des morceaux, douceur, dynamique*.
- Les "interactions" $z_i^m$ entre individus et genres musicaux, par les notes d'appréciation que les premiers décernent aux seconds.

Nous cherchons à analyser les préférences musicales en fonction des caractéristiques individuelles et de celles des genres de musique. Plus précisément, ce sont les différences de *profils de notation* que l'on cherche à modéliser, entre individus comme entre genres. On a donc choisi de modéliser la liaison entre *individu* et *genre musical*, comme si l'on avait affaire à deux variables nominales. Du tableau $Z$ considéré comme un tableau de contingence, on tire le tableau $\boldsymbol{\Phi}$ selon le procédé du § 3.3.*b*. Ce faisant, un poids plus grand est donné aux individus notant plus généreusement, comme aux genres les plus appréciés. On applique FILM-A au tableau $\boldsymbol{\Phi}$.

**a) Analyse ne tenant pas compte de la force structurelle:**

***Calcul des composantes***

Les trois premiers couples de composantes $(f^1,g^1)$, $(f^2,g^2)$, $(f^3,g^3)$ permettent de restituer 76% de la variance des interactions ( $R^2=\|\hat{\Phi}\|^2/\|\Phi\|^2=0.76$ ). Sur le plan de la force structurelle, les composantes captent les parts suivantes de la variance de leur groupe:

$f^1$ : 26% ; $f^2$ : 12% ; $f^3$ : 9%

$g^1$ : 40% ; $g^2$ : 6% ; $g^3$ : 4%

La régression de $\boldsymbol{\Phi}$ sur les $U^{st}=f^s g^t$ fournit le modèle latent estimé:

$$\boldsymbol{\Phi} = 0.75 f^1 g^{1'} + 0.34 f^2 g^{2'} + 0.30 f^3 g^{3'} + E$$

Avec les parts de variance expliquée respectives:

$$f^1 g^{1'} = 56\% \;\; ; \;\; f^2 g^{2'} = 11\% \;\; ; \;\; f^3 g^{3'} = 09\%$$

Le couple de rang 1 est nettement dominant.

### *Interprétation des composantes:*

**Composantes-sujet:**

- Sur la figure 7, il apparaît que la première composante sujet $f^1$ est fortement corrélée négativement: au revenu, au niveau d'études ainsi qu'à l'âge et positivement: au statut d'activité étudiant.
- La deuxième composante sujet $f^2$ est pauvrement illustrée, n'étant bien corrélée, positivement, qu'à l'écoute de musique sur CD.

*Figure 7: Plan sujet (1,2)*
*N.B.:Les 18 sujets sont codés* x1 *à* x18

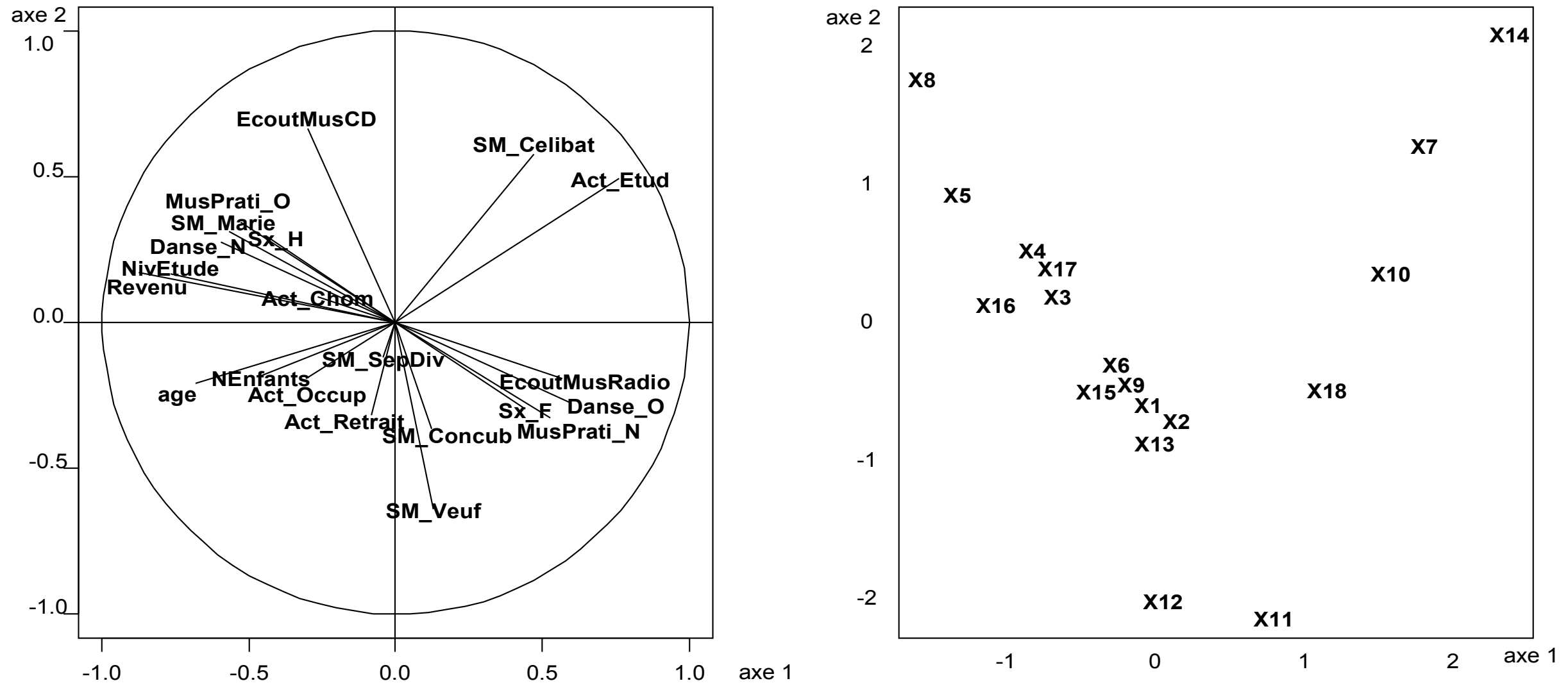

**Composantes-objet:**

- Sur la figure 8, il apparaît que la première composante objet $g^1$ est fortement corrélée (négativement) à un faisceau de variables traduisant la complexité et la richesse de la musique. Ces variables sont anticorrélées à la répétitivité et la dansabilité.
- La deuxième composante objet $g^2$ n'est corrélée à aucune caractéristique musicale.

### *Interprétation des interactions:*

Compte tenu des faiblesses structurelles des composantes $g^2$ et $g^3$, on ne peut retenir dans le modèle que le couple ($f^1$,$g^1$). Bien que restituant plus de la moitié des interactions, ce modèle est conceptuellement relativement maigre: il révèle simplement que maturité et niveau socio-culturel sont associés à une préférence pour les musiques plus complexes et moins dansantes (*opéra, jazz contemporain, musique de chambre, baroque...*).

*Figure 8: Plan objet (1,2)*

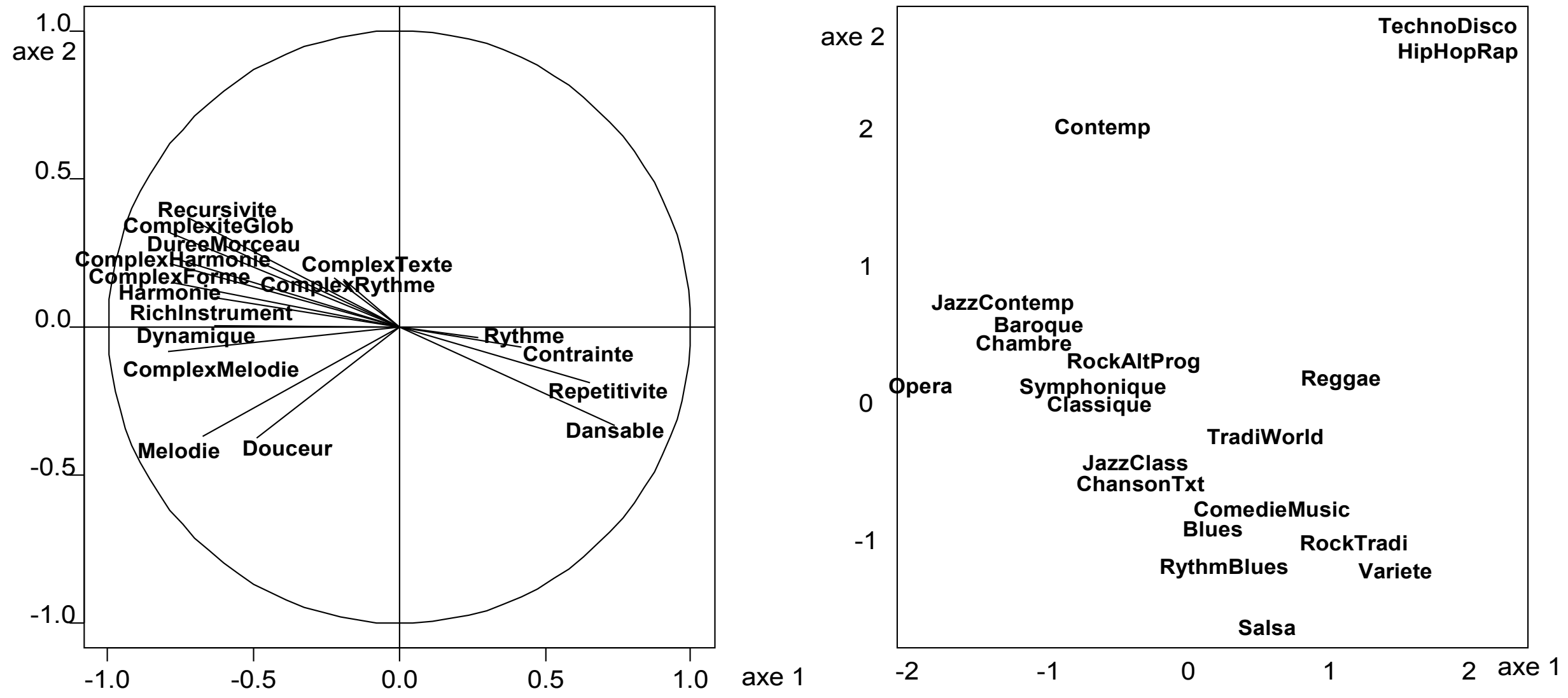

**b) Analyse tenant compte de la force structurelle:**

Les 9 premiers couples de composantes issus de $(f^1, f^2, f^3)$ et $(g^1, g^2, g^3)$ permettent de reconstituer 53% de la variance des interactions, ce qui est inférieur au seul premier couple de l'analyse sans prise en compte de la force structurelle. On espère toutefois que les composantes vont s'avérer plus riches, et permettront ainsi de dépister des phénomènes précédemment passés inaperçus.

Forces structurelles:

$f^1$ : 32% ; $f^2$ : 13% ; $f^3$ : 11%

$g^1$ : 53% ; $g^2$ : 11% ; $g^3$ : 5%

La régression de $\Phi$ sur les $U^{st} = f^s g^t$ fournit le modèle latent estimé:

$$\begin{aligned}\Phi \quad = \quad & \underline{\mathbf{0.61}\, \boldsymbol{f^1 g^{1\prime}}} \quad + \quad \underline{0.12\, f^1 g^{2\prime}} \quad + \quad \mathbf{-0.20}\, \boldsymbol{f^1 g^{3\prime}} \\ + \quad & \underline{0.12\, f^2 g^{1\prime}} \quad + \quad \underline{\mathbf{0.20}\, \boldsymbol{f^2 g^{2\prime}}} \quad + \quad 0.10\, f^2 g^{3\prime} \\ + \quad & -0.14\, f^3 g^{1\prime} \quad + \quad 0.02\, f^3 g^{2\prime} \quad + \quad 0.11\, f^3 g^{3\prime} \quad + \quad E\end{aligned}$$

Nous avons fait figurer en gras les interactions les plus fortes. Signalons qu'avec le seul vecteur $f^1 g^{1\prime}$, on reconstitue 37% de $\Phi$, qu'avec les termes de rang 2: $f^1 g^{1\prime}, f^1 g^{2\prime}, f^2 g^{1\prime}, f^2 g^{2\prime}$, (soulignés), on obtient 44%, et enfin, qu'avec les trois termes d'interaction les plus forts: $f^1 g^{1\prime}, f^2 g^{2\prime}, f^1 g^{3\prime}$ (gras), on obtient 45%.

Le couple de rang 1 est ici encore dominant.

### *Interprétation des composantes:*

**Composantes-sujet:**

La troisième composante n'interagissant guère avec les autres, on se cantonnera aux deux premières (*cf.* figure 9).

- La première composante-sujet conserve grossièrement la même interprétation que précédemment (niveau socio-culturel & maturité): fortement corrélée - positivement cette fois - au revenu et au niveau d'étude. Cependant, elle est ici moins liée à l'âge et davantage à la modalité *Homme* dans le sens du *niveau d'étude* et du *revenu*. D'autre part, on voit apparaître à l'opposé des variables qui lui sont fortement anticorrélées: écoute la musique à la *radio*, *ne pratique pas la musique*, et *Femme*. Cette composante est donc plus "riche".
- La seconde composante n'est toujours pas parfaitement illustrée, mais les variables qui lui sont les plus corrélées sont très bien représentées dans le premier plan. On y voit ainsi apparaître un petit faisceau diagonal, proche de $f^2$, constitué des statuts *célibataire* et *étudiant*, ainsi que de l'*âge*. Cette dimension interagit donc avec celles du premier plan formé par les composantes-objet.

*Figure 9: Plan sujet (1,2)*

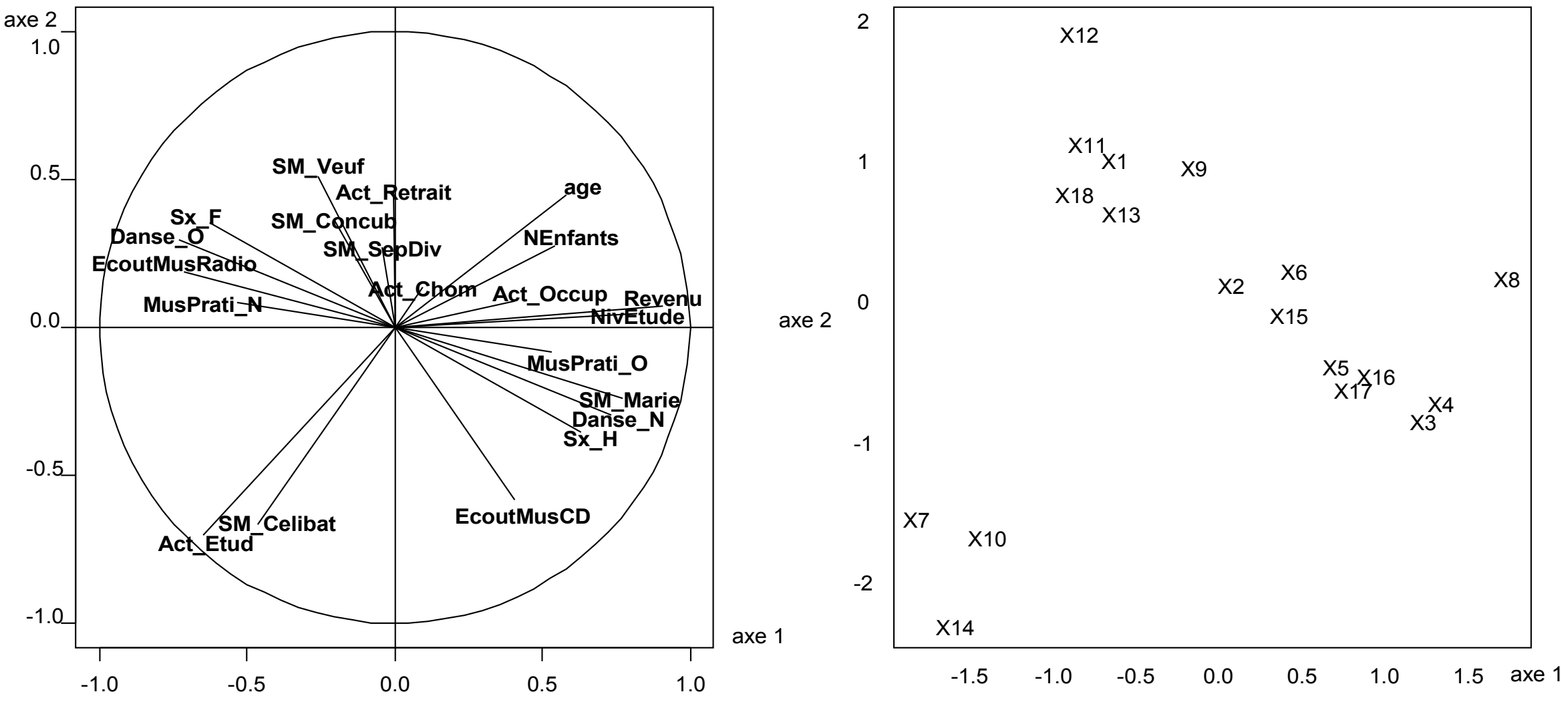

**Composantes-objet:**

La troisième composante $g^3$ interagit ici avec $f^1$; on cherchera donc à interpréter le sous-espace engendré par les trois premières composantes (*cf.* figures 10 et 11).

- La première composante-objet est ici encore essentiellement une composante de *complexité*.
- La seconde composante-objet est très liée à la *douceur* et à l'importance de la *mélodie*.
- La troisième composante-objet est mal illustrée, mais dans le plan ($g^1$,$g^3$), la variable *dansable* est très bien représentée, négativement corrélée à $g^1$ et positivement à $g^3$.

*Figure 10: Plan objet (1,2)*

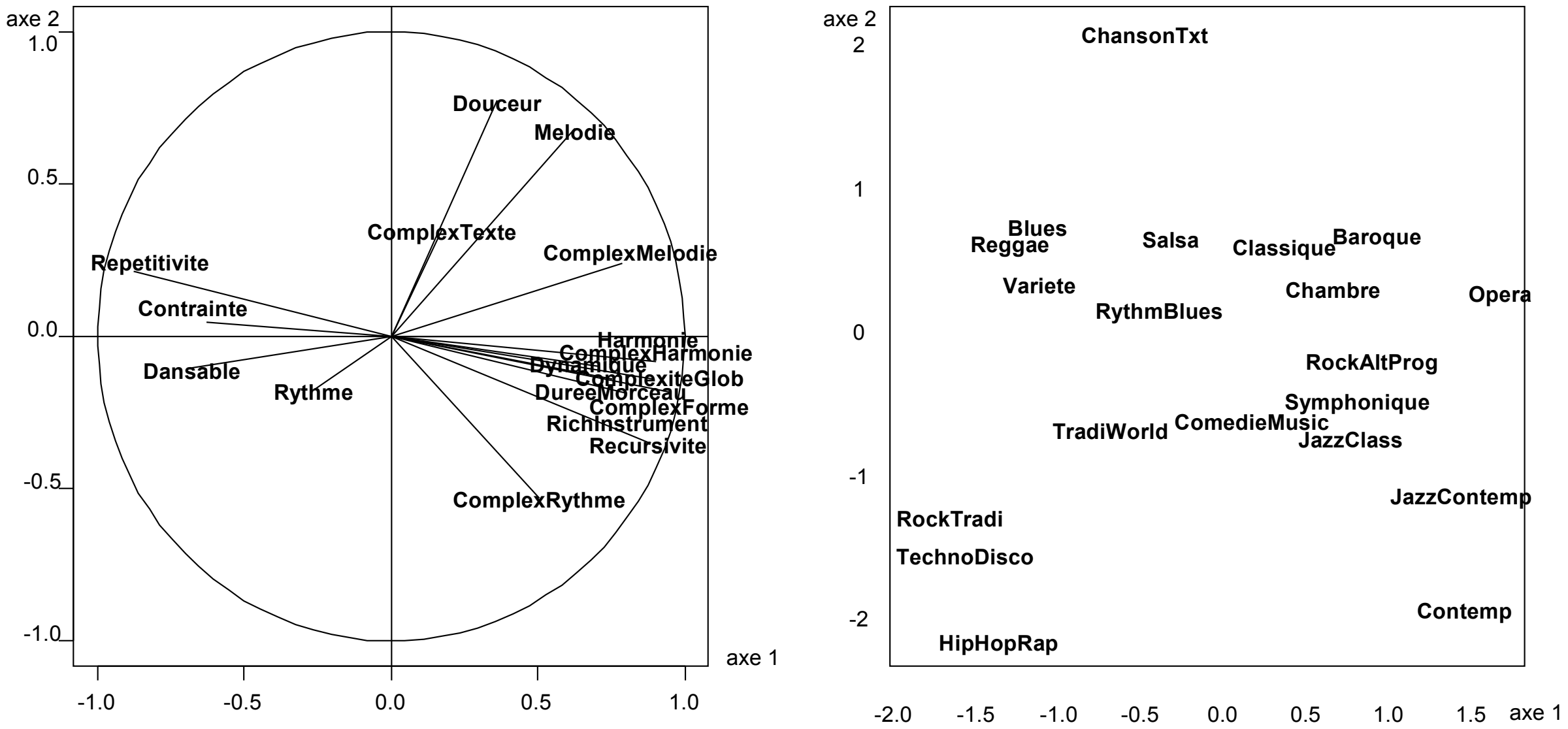

*Figure 11: Plan objet (1,3)*

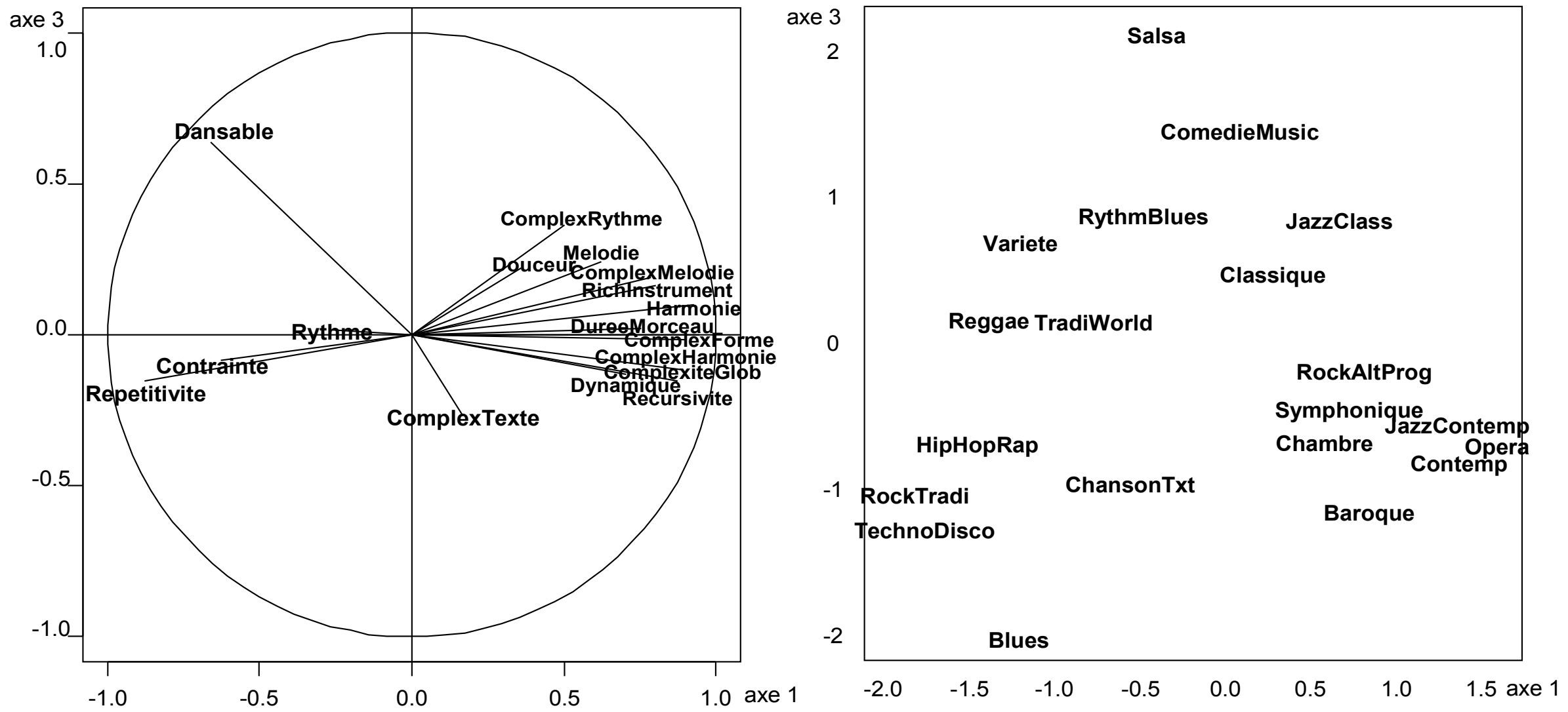

***Interprétation des interactions:***

- L'interaction des composantes sujet et objet de rang 1 est positive et s'interprète comme dans le cas précédent.
- L'interaction des composantes de rang 2 est positive également, et signale que moins on est étudiant/célibataire/jeune, plus on apprécie musiques douces et mélodiques. Les jeunes, en effet, apprécient ici davantage *rock traditionnel*, *techno-disco*, *hip-hop-rap*.
- $g^3$ interagit avec $f^1$, mais n'est pas bien illustrée en tant que telle. Par contre, le plan ($g^1$,$g^3$) fait apparaître *dansable*. On cherche donc à interpréter l'interaction de $f^1$ avec le plan ($g^1$,$g^3$) , en factorisant les termes d'interaction faisant intervenir $f^1$:

$$0.61 f^1 g^{1\prime} - 0.20 f^1 g^{3\prime} = -f^1 (-0.61 g^1 + 0.20 g^3)'$$

Anticorrélée à $f^1$ se trouve la variable-sujet *danse_oui*, et très proche de la combinaison ($- 0.61g^1 + 0.20\, g^3$) se trouve la variable-objet *dansable*, ce qui se passe de commentaire. Mentionnons seulement qu'on peut immédiatement repérer les musiques les plus dansables sur le plan (1,3) des objets: *salsa, variété, reggae, rock traditionnel*, *techno-disco*, *hip-hop-rap*.

**c) Bilan:**

En tenant compte de la force structurelle des composantes, on a certes obtenu un modèle moins parcimonieux, moins bien ajusté, mais beaucoup plus riche d'interprétation: il a permis de dépister 3 phénomènes d'interaction au lieu d'un.

# Conclusion

La méthode FILM est, comme la régression PLS multivariée, au confluent des méthodes factorielles, exploratoires, et des méthodes de régression, modélisantes. Comme PLS, elle exploite un critère de covariance.

La régression PLS est fondée sur le critère de covariance de l'Analyse Inter-Batteries (purement exploratoire), mais en itérant la maximisation de ce critère et le calcul de résidus de régression, elle permet de modéliser l'un des deux tableaux par l'autre. De façon similaire, dans sa version de base, FILM utilise itérativement la maximisation du même critère que les analyses RLQ et L-PLS ainsi que le calcul de résidus de régression pour modéliser le tableau des interactions à partir des descriptions de ses marges.

On peut aussi voir les choses sous l'angle suivant: PLS, en utilisant un critère de covariance au lieu d'un critère de corrélation, greffe la question de la force structurelle des prédicteurs sur la régression linéaire. De la même façon, FILM greffe la force structurelle des prédicteurs sur le problème d'ajustement du tableau $Z$ par un modèle d'interactions entre descriptions marginales.

FILM, comme L-PLS, étend ainsi la régression PLS aux données d'interaction, plus complexes. L'usage du critère de covariance lui confère les qualités des méthodes factorielles descriptives - puissance de synthèse à travers une hiérarchie de dimensions fortes, représentations planes des structures multidimensionnelles, conduisant à une analyse simultanée des nuages des observations et des variables. En outre, étant fondée sur un modèle explicatif, elle met directement ces qualités au service de la modélisation. Contrairement à L-PLS, FILM utilise à chaque étape l'ensemble des interactions possibles entre composantes sujet et objet.

# Remerciements

Nous remercions vivement Pierre Cazes pour sa lecture attentive et éclairée, ainsi que Robert Sabatier et Eric Gorouben pour leurs conseils.

# Bibliographie